\documentclass[draft]{agujournal2019} 
\usepackage{url} 
\usepackage{lineno}
\usepackage{soul}
\usepackage[english]{babel}
\usepackage{float}
\usepackage{apacite}
\usepackage{multirow}
\usepackage{array}
\usepackage{siunitx}
\usepackage{amsmath}
\usepackage{amssymb}
\usepackage{bibentry} 
\usepackage{appendix}

\draftfalse
\journalname{Journal of Geophysical Research - Planets}
\graphicspath{{Figures/}}

\newcommand{\degree}{$^\circ$}
\newcommand{\dragE}{~m s$^{-1}$ day$^{-1}$}
\newcommand{\drag}{~m s$^{-2}$}

\newcommand{\flux}{~m$^2$ s$^{-2}$}

\soulregister\flux7
\soulregister\degree7
\soulregister\drag7
\soulregister\dragE7

\usepackage{booktabs}
\begin{document}
\title{Variation of Venusian Gravity Wave Absolute Momentum Fluxes and Drag as Retrieved from the Akatsuki Mission}
\authors{Erdal Yi\u git \affil{1,2}, Emilia Sloan\affil{3}}

\affiliation{1}{Department of Physics and Astronomy, Space Weather Lab, George Mason University, Fairfax, VA, USA}
\affiliation{2}{Ionosphere Thermosphere Mesosphere Lab (Code 675), NASA Goddard Space Flight Center, Greenbelt, MD, USA}
\affiliation{3}{Thomas Jefferson High School for Science and Technology, Alexandria, VA, USA}
\correspondingauthor{Erdal Yi\u git}{eyigit@gmu.edu}


\begin{keypoints}
\item Middle atmospheric gravity wave fluxes and drag are retrieved from the Akatsuki mission. 
\item Altitude variation of gravity wave fluxes demonstrate the effects of wave damping most likely due to nonlinear dissipation of gravity waves.
\item  Maximum absolute gravity wave horizontal momentum flux and drag vary between 10--30\flux~ and 0.003--0.03\drag, depending on the latitude. 
\end{keypoints}


\begin{abstract}
Using temperature retrievals from Akatsuki radio occultation measurements, we characterize gravity wave activity as a function of vertical wavenumber and altitude and, for the first time, estimate the absolute horizontal momentum fluxes and the magnitude of the associated gravity wave drag (i.e., wave acceleration), which quantify the potential effects of these waves in the Venusian middle atmosphere between 40--95 km. Observed temperature perturbations, which are indicative of atmospheric gravity wave activity, reach amplitudes of approximately $\pm$10 K, and significant momentum flux (10--30\flux) and wave drag (0.003--0.03\drag) are detected across all analyzed profiles. The inferred wave drag represents a lower bound on the total gravity wave-induced drag in the Venusian atmosphere. Momentum flux tends to increase exponentially with altitude below approximately 50--60 km, then peaks and attenuates at higher altitudes. Wave drag becomes prominent where momentum flux begins to decrease, which is a consequence of wave dissipation. Both quantities exhibit multiple altitude-localized maxima, which is consistent with upward wave propagation followed by dissipation at different altitudes for different vertical wavelengths. Damping due to gravity wave nonlinear interactions is likely to play the major role in limiting the growth of wave amplitudes and fluxes with height.  
These features are observed across a range of latitudes and local times. Overall, the results provide observational constraints on gravity wave momentum transport and dissipation in the Venusian middle atmosphere and could guide numerical models in their effort to quantify wave-mean flow interactions in Venus's atmosphere. 
\end{abstract}

\section*{Plain Language Summary}
Atmospheric gravity waves are present in all planetary atmospheres, including Venus’. We use temperature observations from the Akatsuki satellite to detect and to estimate gravity wave-induced momentum flux and wave drag, which have not previously been estimated from observations for Venus. Momentum flux describes how gravity waves transport momentum upward through the atmosphere. Wave drag describes how gravity waves can influence the background winds. Our estimate represents the strength of this effect but does not indicate whether winds are sped up or slowed down. We found that momentum flux and wave drag are present throughout Venus’s atmosphere and show similar patterns across latitudes and local times, including an initial increase in momentum flux with height. These results improve our understanding of how gravity waves behave in Venus’s middle atmosphere. 

\section{Introduction}

Internal gravity waves are produced by the restoring action of buoyancy, which results from gravity acting on density stratification, in geophysical and astrophysical fluids. They are ubiquitous features of all stably stratified planetary atmospheres \cite{Yigit.Medvedev2019_ObscureWaves}. They have been extensively observed in all altitudes on Earth \cite{Nyassor.etal2025_MomentumFlux, Forbes.etal2016_GravityWaveinduced} and Mars \cite{Imamura.etal2024_ShortVertical, Ando.etal2012_VerticalWavenumber, Heavens.etal2020_MultiannualRecord, Yigit.etal2015_HighaltitudeGravity, England.etal2017_MAVENNGIMS, Jesch.etal2019_DensityFluctuations} and shape the dynamical and thermal structure of the whole atmosphere region on Earth and Mars \cite{Yigit.etal2025_ImpactGravity, Liu.etal2025_DiurnalCycle} and produce tracer transport \cite{Shaposhnikov.etal2022_MartianDust,Liu.etal2025_StochasticParameterization}.


 Small-scale atmospheric waves have been identified in the Venusian atmosphere from cloud top images \cite{peralta_characterization_2008, Titov.etal2012_MorphologyCloud, Piccialli.etal2014_HighLatitudea, silva_characterising_2021, silva_atmospheric_2024},  from temperature fluctuations \cite<e.g.,>[]{tellmann_small-scale_2012, ando_vertical_2015}, on the dayside CO$_2$ non-LTE emissions above the clouds \cite{Garcia.etal2009_GravityWaves}, in vertical profiles measured during the descent of entry probes of space missions like Pioneer Venus and Veneras \cite{Seiff.etal1992_EvidencesWaves}, as well as in vertical profiles of the oxygen nightglow obtained with the Visible and Infrared Thermal Imaging Spectrometer (VIRTIS) onboard Venus Express \cite{Altieri.etal2014_ModelingVIRTIS}. Waves have been identified not only on the ultraviolet albedo of the dayside top of the clouds, but also on the albedo of the middle clouds \cite{Peralta.etal2008_CharacterizationMesoscale}, the opacity of the nightside lower clouds \cite{Silva.etal2021_CharacterisingAtmospheric, Peralta.etal2008_CharacterizationMesoscale}, and on the brightness temperature of the day/nightside upper clouds \cite{Peralta.etal2017_StationaryWaves, Kouyama.etal2017_TopographicalLocal}. Across the Venus Express and Akatsuki missions, comparisons with (linear) wave saturation theory find some agreement between the observed vertical wavenumber ($m$) spectra of temperature perturbations and the expected $m^{-3}$ power spectrum above 60--65 km altitude \cite{ando_vertical_2015, mori_gravity_2021, pramitha_vertical_2025}. The agreement has been interpreted as the result of superposition of unsaturated quasi-monochromatic wave packets \cite{mori_gravity_2021} or spectra composed of waves approaching breaking, and attenuated by turbulence \cite{ando_vertical_2015}. The vast majority of these observations interpret the retrieved wave-like small-scale structures as atmospheric gravity waves. 

Internal atmospheric waves transfer energy and momentum to the background atmosphere via critical level filtering, convective instability, and mechanical dissipation, which can be due to radiative damping, nonlinear wave-wave interactions, molecular viscosity and thermal conduction, and ion-neutral friction \cite{Yigit.etal2008_ParameterizationEffects}. Critical filtering of gravity waves by the background atmosphere occurs when the speed and the direction of the gravity wave and the background wind are equal to each other at a given altitude, i.e., when the wave intrinsic horizontal phase speed is zero, $\hat{c}_i = |c_i-\bar{u}|= 0 $, where $\bar{u}$ is the background mean wind and $c_i$ is the horizontal phase speed of the probe wave "i".  Gravity waves experience convective instability when the wave-induced fluctuation (or just the wave amplitude) $|u^\prime_i|$ exceeds the instability threshold, given by the absolute value of the intrinsic phase speed of the wave, at the so-called wave breaking level:  $|u_i^\prime| \ge |c_i-\bar{u}|$. Radiative damping has been considered on Venus as a potential sources of wave dissipation, which can shape the observed vertical wavenumber spectra of gravity waves \cite{ando_vertical_2015, mori_gravity_2021}. Accordingly, longer vertical wavelength waves are expected to amplify with altitude and approach saturation, whereas shorter vertical wavelength waves could exhibit negative growth and weaken with altitude \cite{ando_vertical_2015}, matching the observed decline in short wavelength waves \cite{mori_gravity_2021} and a decrease in temperature perturbation amplitudes with altitude \cite{tellmann_small-scale_2012}. Radiative damping-based frameworks have been used to interpret gravity wave observations, while multiple studies suggest that wind shear and the mean zonal flow could significantly impact gravity wave propagation and inferred characteristics \cite{ando_vertical_2015, Noguchi.etal2025_RadioScintillation}. The other dissipative processes mentioned above are less explored on Venus, however, extensive global-scale numerical studies on Earth and Mars demonstrate that nonlinear diffusion is the primary source of wave dissipation in terrestrial middle atmospheres, which molecular diffusion takes over at greater heights in the thermosphere as the dominant source of wave dissipation \cite{Medvedev.etal2011_EstimatesGravity, Yigit.etal2009_ModelingEffects, Yigit.etal2012_DynamicalEffects, Gavrilov.Kshevetskii2015_DynamicalThermal}. In nonlinear interactions of gravity waves the low-frequency (higher vertical wavenumber) part of the gravity wave spectrum acts as an additional background to the high-frequency waves. Specifically, gravity wave harmonics with wavelengths (and periods) longer than that of the given wave serve as an additional background and induce a "nonlinear" Doppler shift \cite{Weinstock1975_NonlinearTheory, Weinstock1982_NonlinearTheory, Medvedev.Klaassen1995_VerticalEvolution, Yigit.etal2008_ParameterizationEffects}.

The spatial and temporal distribution of gravity wave activity is used to understand potential sources of gravity waves in the Venusian atmosphere. Multiple studies have found that gravity wave activity tends to increase near polar latitudes \cite<e.g.,>[]{tellmann_small-scale_2012, ando_vertical_2015, mori_gravity_2021, pramitha_vertical_2025}. The increase in activity can be attributed to increased convection in the middle cloud at polar latitudes \cite{tellmann_small-scale_2012}. Other studies propose cloud-level convection as a gravity-wave source that could explain stronger high-latitude activity \cite<e.g.,>[]{leroy_convective_1995, leroy_radio_1996, baker_convectively_2000, baker_convectively_2000-1, imamura_inverse_2014, peralta_characterization_2008, tellmann_small-scale_2012, ando_vertical_2015, lefevre_threedimensional_2017, lefevre_threedimensional_2018}. Multiple studies using visual observations of wave packets in the clouds from cameras on Akatsuki and Venus Express have not identified significant differences in the characteristics of gravity wave at different latitudes \cite<e.g.,>[]{peralta_characterization_2008, silva_characterising_2021, silva_atmospheric_2024, espadinha_wave_2026}, which has been extended to suggest deep convection as a source \cite{peralta_characterization_2008}. Sources such as shear/instabilities associated with the mid-latitude jet and thermal-tide jet-exit regions have also been proposed \cite{tellmann_small-scale_2012, imamura_momentum_1997, sugimoto_generation_2021}. Multiple observational studies find little or no local time dependence of gravity-wave activity on Venus \cite{peralta_characterization_2008, imamura_inverse_2014, Noguchi.etal2025_RadioScintillation}, whereas \citeA{tellmann_small-scale_2012} report a local time variation attributed to day/night differences in convective activity. 

While spatial and temporal distributions of gravity wave activity help to identify potential sources of excitation, more direct retrieval of gravity wave momentum and energy transport are essential for assessing their overall influence on Venus’s atmosphere. Internal gravity waves produce vertical flux of horizontal momentum from their sources in the lower atmosphere to higher altitudes. The divergence of the wave momentum flux produces irreversible momentum deposition to the background atmosphere, resulting in the acceleration/deceleration of the neutral winds \cite{Andrews.etal1987_MiddleAtmosphere,Yigit.Medvedev2015_InternalWave, Medvedev.Yigit2019_GravityWaves}. Therefore, quantification of the vertical flux of the horizontal gravity wave momentum is essential to our understanding of the Venusian atmospheric circulation at all altitudes. General circulation models coupled to gravity wave parameterizations require gravity wave momentum fluxes both as an input at wave sources and as a tool for validation \cite{Medvedev.Yigit2019_GravityWaves}. Modeling studies using parameterized momentum fluxes suggest that gravity wave drag can significantly shape Venusian winds, from the cloud-top to the thermosphere, where the produced drag can impact the thermal tides \cite{Zalucha.etal2013_IncorporationGravity, sugimoto_generation_2021}. Gravity wave drag is essentially a proxy for the dynamical effects on the mean flow due to dissipating gravity waves. 

We use radio occultation data from the Akatsuki mission \cite{murakami_venus_2017} to estimate, for the first time, the absolute horizontal momentum flux and the associated gravity wave drag in the Venusian atmosphere as a function of vertical wavelength and altitude. We examine the wave propagation across wavelengths, providing potential insights into amplitude growth with altitude and wave dissipation. We further investigate the latitude and local-time variation of these parameters to provide an idea of the global distribution of the momentum transport by gravity waves and discuss common features in the momentum flux and drag profiles.

We begin with an overview of the Akatsuki mission and the temperature profiles used in this study (Section \ref{sec:akatsuki}). We then describe how gravity wave perturbations are isolated and used to estimate momentum flux and wave drag, along with assumptions and limitations of the approach (Section \ref{sec:gw_retrieval}). We present momentum flux and drag spectra and their variation with altitude, vertical wavelength, latitude, and local time in Section \ref{sec:results}. We interpret these results in terms of wave growth and dissipation and potential source regions to contextualize the inferred flux and drag magnitudes in terms of the larger atmospheric circulation (Section \ref{sec:discussion}). Summary and conclusions are given in Section \ref{sec:summary}.

\section{Data \& Methods}
\label{sec:data}

\subsection{Akatsuki Mission}
\label{sec:akatsuki}
The Akatsuki Mission \cite{Nakamura.etal2011_OverviewVenus, nakamura_akatsuki_2016}, launched on March 21, 2010 by the Japan Aerospace Exploration Agency (JAXA), successfully surveyed the Venusian atmosphere and its climate from December 2015 till late-April 2024, when contact was lost to the spacecraft. Akatsuki's Radio Science (RS) instrument conducts radio occultation experiments by transmitting one-way X-band downlink signals. These radio signals propagate through Venus's atmosphere and refract before being received at the Usuda Deep Space Center (USDC) or the Indian Deep Space Network (IDSN). The refractive index is determined as a function of altitude using the received signals. The number density profiles of the neutral atmosphere and the ionosphere are retrieved from the refractivity profiles. The hydrostatic equilibrium equation and the ideal gas law are applied to the density measurements to retrieve atmospheric pressure and temperature \cite{imamura_initial_2017}. We use the retrieved Level 4 vertical profiles of temperature, pressure, and density archived in the Data ARchives and Transmission System (DARTS) \cite{murakami_venus_2017}.

\begin{figure}
 \begin{centering}
    \includegraphics[width=1.0\linewidth, keepaspectratio]{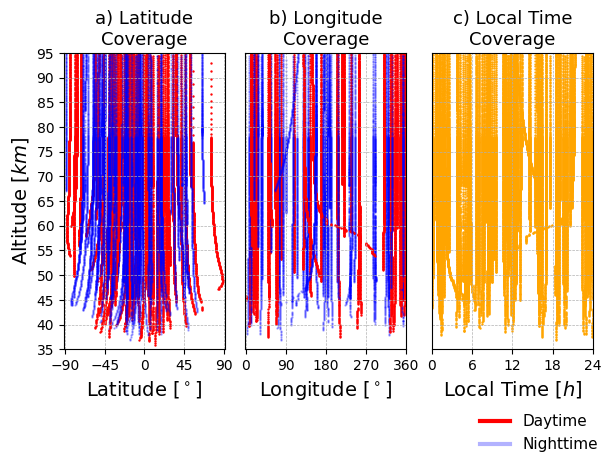}
    \caption{{Each of these panels plots data from all radio occultation temperature profiles used. (a) Latitude is plotted on the x-axis, altitude on the y-axis for each profile. (b) Longitude is plotted on the x-axis, altitude on the y-axis for each profile. Profiles plotted in red represent daytime observations, profiles plotted in blue represent nighttime observations. (c) Local time is plotted on the x-axis, altitude in the y-axis for each profile.}}
    \label{fig:fig1_new}
  \end{centering}
\end{figure}

We present the coverage of latitude, longitude, and local time by examining the variation of each of these variables with altitude for each profile in Figure \ref{fig:fig1_new}. Akatsuki’s near-equatorial orbit provides good coverage of low and midlatitudes in both hemispheres; coverage of high latitudes is more limited. Data gaps are present in the longitude and local time coverage, but overall, most longitudes and local times are covered across the 138 radio occultation profiles used for this study. More profiles might be available, however, the result of Akatsuki’s failed initial insertion into orbit was a decrease in the frequency of radio occultations due to the elongation of Akatsuki's orbit from $\sim 30$ hours to $\sim 10.5$ days. Additionally, radio occultation measurements require the spacecraft to be occulted by Venus as viewed from a tracking station of Earth. Radio occultations only occur when this geometry is achieved, which is less frequently than every orbit \cite{nakamura_akatsuki_2016}.

\subsection{Method of Gravity Wave Absolute Momentum Flux Retrieval}
\label{sec:gw_retrieval}
We use 138 temperature-altitude profiles to first calculate temperature perturbations. The profiles used are collected between 2016--2024. The method we use to determine the absolute momentum flux and wave drag is based on the approach described in the works by \citeA{Starichenko.etal2021_GravityWave} and \citeA{Ern.etal2004_AbsoluteValues}. Nevertheless, we will outline here the relevant details of our retrieval method. The observed temperatures $T$ can be decomposed into a mean background temperature $\bar{T}$, and the fluctuation term $T^\prime$ as $T = \bar{T} + T^\prime$, where the fluctuation term is representative of gravity wave activity resulting from the superposition of various gravity wave spatiotemporal scales. To extract $\bar{T}$ and $T^\prime$ from temperature-altitude profiles, the temperature is first interpolated onto an altitude grid intended to create an even distance between consecutive altitude levels. The altitude spacing is set to 0.2 km, which is similar to the average altitude spacing of the data. After interpolation, a 7th-degree polynomial is fitted to temperature as a function of altitude for the entire profile to determine $\overline{T}$, as done in numerous studies of gravity waves in planetary atmospheres \cite{Yigit.etal2015_HighaltitudeGravity, Yigit.etal2021_VariationsMartian, Starichenko.etal2021_GravityWave, Leelavathi.Rao2024_ComparativeAnalysis}. $T^\prime$ is determined by subtracting $\overline{T}$ from the observed temperature at each altitude in the profile: $T^\prime = T - \bar{T}$.

The temperature perturbations are then separated into vertical wavelength ($\lambda_z$) windows. The chosen window length is 15 km, meaning that the maximum vertical wavelength, that can be resolved is 15 km. The windows are rectangular. Starting at the altitude of the first measurement, temperature perturbations are taken from 7.5 km above and below each altitude. Beginning at the top and bottom of the profiles, the windows are shifted by 1 km until the entire profile is separated into windows. The profiles that are separated into windows beginning at the bottom of the profile are called upward, and the profiles that are separated beginning at the top of the profile are called downward.

The Fourier transform is applied to each 15 km window with the vertical wavenumber, $m$, ranging from 0.067 cycles km$^{-1}$ to $2.5$ cycles km$^{-1}$. Each value of $m$ is considered a wave harmonic. For each window, the power is calculated for each $m$ or each harmonic as $P=\frac{2}{N_{s}^2}|X|^2$, where $X$ values are the Fourier coefficients and $N_{s}$ is the window length corresponding to the total number of data points within a window, which, with a 15 km window and 0.2 km spacing, is 75 data points. The factor of 2 results from the use of a one-sided Fourier transform. The power per harmonic is assigned to the altitude at which each window is centered.

After applying the Fourier transform to all windows for the upward and downward profiles, the power is averaged for common altitudes. The amplitude, given by $A =\frac{2}{N_s}|X| = \sqrt{2P}$, is recovered from the averaged powers. Since there is rarely an integer number of steps between the minimum and maximum altitudes, amplitudes are averaged when an altitude level in the downward and upward profiles are within 0.5 km of each other. This only occurs once for each altitude level due to the 1 km step used when windowing the data. The averaged amplitudes are then assigned to the averaged altitudes from the upward and downward profiles. The maximum and minimum altitudes are excluded since they are not within 0.5 km of an altitude from the opposing upward/downward profile. The Fourier transform, upward and downward averaging, and amplitude retrieval were tested on multiple synthetic waves with known wavelengths and amplitudes.

The Brunt-Väisälä frequency (or buoyancy frequency), $N$, is calculated as a function of altitude. Gravity is taken as altitude-dependent, calculated using the mass ($\sim$ 4.87 $\times 10^{24}$ kg) and radius ($\sim 6051.8$ km) of Venus recorded in \citeA{archinal_report_2018} and NASA's Planetary Physical Parameters. Between 35--95 km, the acceleration decreases from $\sim 8.77$ m s$^{-2}$ to $\sim 8.60$ m s$^{-2}$. The specific heat capacity at constant pressure, $c_p = 850.1$ J K$^{-1}$ kg$^{-1}$, is taken as a constant, assuming a CO$_2$ only atmosphere. We exclude the regions of profiles with $N^2 < 1 \times 10^{-5}$ to prevent an nonphysical increase in the absolute gravity wave momentum flux, $F_{k,m}$, where $k$ and $m$ are the horizontal and vertical wavenumbers, respectively. The Brunt-Väisälä frequency and the absolute gravity wave momentum flux are given by:

\begin{linenomath*}
\begin{equation}
\label{eq:buoyancy-frequency}
 N = \sqrt{\frac{g}{\overline{T}}\bigg(\frac{d\overline{T}}{dz}+\frac{g}{c_p}\bigg)},  
\end{equation}
\end{linenomath*}

\begin{linenomath*}
\begin{equation}
\label{eq:absoluteflux}
 F_{k,m} = \frac{1}{2}\frac{k_h}{m}\bigg(\frac{g}{N}\bigg)^2 \bigg(\frac{|T^\prime_{k,m}|}{\overline{T}} \bigg)^2,   
\end{equation}
\end{linenomath*}
where $k_h = 2\pi/\lambda_h$ is the horizontal wavenumber with the horizontal wavelength $\lambda_h$ and $T^\prime_{k,m}$ is the amplitude of the harmonic with a given horizontal and vertical wavenumber. Multiplying \eqref{eq:absoluteflux} with the height-dependent background density $\rho$ yields wave-induced stress. In \eqref{eq:absoluteflux} the only free parameter is the horizontal wavenumber $k_h$, while others are provided by or retrieved from observations. For this, we use a representative horizontal wavelength of $\lambda_h = 300$ km, which is a standard value often used in studies of gravity waves in terrestrial planets, such as Earth, Mars, and Venus \cite{Yigit.etal2009_ModelingEffects, Yigit.etal2018_InfluenceGravity, Hinson.Jenkins1995_MagellanRadio, Yigit.etal2021_EffectsLatitudeDependent, Zalucha.etal2013_IncorporationGravity, Noguchi.etal2025_RadioScintillation}. The consequence of this choice and the associated errors are discussed in \ref{app:uncertainties}. Using the temperature perturbation amplitudes the momentum flux for each harmonic, $F_{k,m}$, is calculated. The total absolute gravity wave horizontal momentum flux is calculated by summing $F_{k,m}$ across all harmonics, $F = \sum_mF_{k,m}$, noting that $k=k_h$ is fixed in our case.

From the total absolute gravity wave horizontal momentum flux, we calculate the absolute horizontal gravity wave drag, which is the wave-induced acceleration (or deceleration) of the mean flow:
\begin{linenomath*}
\begin{equation}
\label{eq:gravity_wave_drag}
   a_h = 
    -\frac{1}{\rho(z)} \frac{\partial [\rho(z) F(z)]}{\partial z} \geq 0,
\end{equation}
\end{linenomath*}
where $\rho$ is the background (mean) density. \eqref{eq:gravity_wave_drag} informs us on the magnitude of the gravity wave drag due to the divergence of the wave stress (or flux) without the directional knowledge. Since density is strictly positive, the term $\frac{\partial}{\partial z}[\rho(z)F(z)]$ determines the sign of the wave drag. Under conservative propagation or dissipation, the absolute wave drag cannot be negative and, therefore, $\frac{\partial}{\partial z}[\rho(z)F(z)]$ cannot be positive. To ensure this term is not positive, we impose the constraint:
\begin{linenomath*}
    \begin{equation}
    \label{eq: flux_inequality}
        \frac{d}{d z}[F(z)\rho(z)] \leq 0,
    \end{equation}
\end{linenomath*}
which can be discretized using a forward finite difference over the altitude interval [$z_k, z_{k+1}$]:
\begin{linenomath*}
    \begin{equation}
    \label{eq: discretized}
        \frac{F_{k+1}\rho_{k+1} - F_k\rho_k}{z_{k+1}-z_{k}} \leq 0.
    \end{equation}
\end{linenomath*}
Then, \eqref{eq: discretized} reduces to the condition:
\begin{linenomath*}
    \begin{equation}
        \label{eq: final_inequality}
        F_{k+1}\rho_{k+1} \leq F_k\rho_k.
    \end{equation}
\end{linenomath*}
When the inequality is violated, the momentum flux profile implies a negative wave drag. To correct such cases, we conserve the momentum flux between adjacent altitude layers while correcting the quantity to satisfy:

\begin{linenomath*}
    \begin{equation}
        \label{eq:correction_equality}
    \tilde{F}_{k+1}\rho_{k+1} = \tilde{F}_k\rho_k.
\end{equation}
\end{linenomath*}
The momentum flux conservation between altitude layers is given by:
\begin{linenomath*}
    \begin{equation}
        \label{eq:conservation}
        \tilde{F}_{k} + \tilde{F}_{k+1} = F_{k} + F_{k+1}.
    \end{equation}
\end{linenomath*}
Here $F_k$ and $F_{k+1}$ denote the (uncorrected) momentum flux estimates at adjacent altitude levels $z_k$ and $z_{k+1}$.  $\tilde{F}_k$ and $\tilde{F}_{k+1}$ denote the adjusted fluxes that satisfy the conservative/dissipative propagation constraint. Using \eqref{eq:correction_equality} and \eqref{eq:conservation}, the equations below are derived for the corrected momentum fluxes within an altitude layer.
 \begin{linenomath*}
\begin{equation}
    \tilde{F}_{k} = \frac{\rho_{k+1}}{\rho_k +\rho_{k+1}}[F_k + F_{k+1}],
\end{equation}
\end{linenomath*}
\begin{linenomath*}
    \begin{equation}
        \tilde{F}_{k+1} = \frac{\rho_k}{\rho_k + \rho_{k+1}}[F_k + F_{k+1}].
    \end{equation}
\end{linenomath*}

We use the relations above to iteratively correct the momentum flux until the values converge, ensuring realistic values of $a_h$ given our assumptions. This correction is based on the approach in the work by \citeA{Brown.etal2022_EvidenceGravity}.

\section{Results}
\label{sec:results}

\begin{figure}
 \begin{centering}
    \includegraphics[width=1.0\linewidth, keepaspectratio]{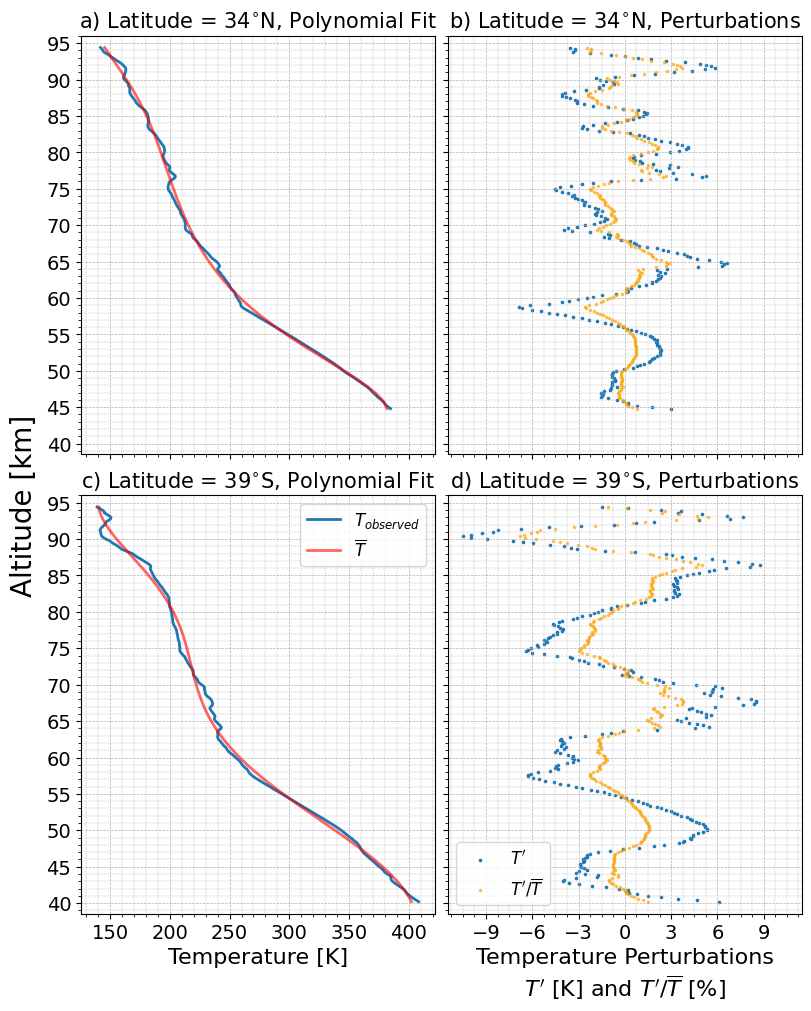}
    \caption{{The top panel is an ingress observation during orbit 216, with latitudes ranging from $32.26^\circ$--$35.60^\circ$N, longitudes ranging from $196.99$--$197.05^\circ$, and solar zenith angle ranging from $42.626^\circ$-- $44.754^\circ$. The bottom panel is an ingress observation during orbit 118, with latitudes ranging from $32.92 - 45.12^\circ$S, longitudes ranging from $67.48^\circ$--$71.23^\circ$, and solar zenith angle ranging from $22.121$-- $32.317^\circ$. (a, c) Observed temperature is plotted in blue. The background temperature, $\overline{T}$, is plotted in red. (b, d) The temperature perturbations, $T^\prime$, are plotted in blue. Normalized temperature perturbations, $\frac{T^\prime}{\overline{T}}$, are shown in orange.}}
    \label{fig:fig1}
  \end{centering}
\end{figure}

Atmospheric gravity waves are typically composed of a spectrum of harmonics with different wavelengths, amplitudes, and fluxes distributed over a wide range of altitudes, exhibiting temporal variability. Here we concentrate on the analysis of gravity-induced temperature amplitudes, absolute horizontal momentum fluxes, and gravity wave drag. For this, we first show for two temperature profiles, one in the northern and one in the southern hemisphere, the application of the method to extract temperature perturbations and other atmospheric parameters needed to calculate momentum flux from occultation profiles. Figure \ref{fig:fig1} presents at two representative middle-latitudes in the northern (34\degree N, upper panels, ingress orbit 216 on June 12 2019) and southern hemispheres (39\degree S, lower panels, ingress orbit 118 on May 27 2022) from $\sim$40--90 km the altitude profiles of background atmospheric and gravity wave related parameters: temperature $T$,  the background mean temperature $\bar{T}$, the gravity wave-induced temperature fluctuation $T^\prime$, and the relative temperature fluctuation $T^\prime/\bar{T}$ due to gravity waves. The temperature decreases with altitude from $\sim$400 K to $\sim 150$ K, which was also documented in the work by \citeA{Ando.etal2020_ThermalStructure}. On the other hand, the overall magnitude of the gravity-wave induced temperature fluctuations and the relative temperature fluctuations generally increase from a few K ($\pm 2$--$3\%$) at 40--50 km up to $\pm 10$ K ($\pm$6--7\%), though the amplitude increase is more conspicuous in the southern mid-latitude profile than in the northern. The increase with altitude in the normalized fluctuations is more pronounced than the 
increase in instantaneous fluctuations, indicating the increasing relative importance of gravity wave effects with altitude. The peak gravity wave activity is generally found at around 60 km in the northern hemisphere profile and between 80--90 km in the southern hemisphere profile. It is possible that the maximum perturbation at the top of the southern hemisphere profile is erroneous due to decreased accuracy of the fit at the edges of the profile. 

Figure \ref{fig:fig2} presents for the two latitudes shown in Figure \ref{fig:fig1} the buoyancy frequency squared $N^2$, $\rho_0/\rho(z)$, and $|T^\prime_m|$ for each spectral band. The buoyancy frequency \eqref{eq:buoyancy-frequency} is an important parameter that informs us on the stability of the atmosphere and is required for the calculation of the gravity wave momentum flux in \eqref{eq:absoluteflux}. It generally increases with altitude varying between $N^2\sim 0.8$--$4\times10^{-4}$ s$^{-2}$ but it has a local minimum around 50 km in both mid-latitude profiles. The ratio of the reference density at around 40 km to the height-dependent-density, $\rho_0/\rho(z)$, demonstrates the decreasing atmospheric density with altitude, which is included due to its relevance to GW amplitude growth with altitude. In each hemisphere it increases by more than 4--5 orders of magnitude from $\sim 40$ km to $ \sim 95$ km, showing a steeper increase at low altitudes compared to higher altitudes. These profiles are not identical but their similarities show that gravity wave amplitudes would increase exponentially with altitude. The spectral amplitude of temperature perturbations, $|T^\prime_m|$, shown as a function of altitude and vertical wavelength $\lambda_z=2\pi/m$ indicates that the longer vertical wavelength waves possess greater amplitudes. Due to the separation of the temperature perturbations into harmonics for each altitude, the altitude where the maximum amplitude in a harmonic occurs does not always reflect the altitude where the temperature perturbations maximize. For example, in the northern hemisphere profile, the amplitude in the first vertical wavenumber bin maximizes at $\sim$ 64 km (Figure \ref{fig:fig1}) whereas the absolute flux in the m = 1 bin maximizes at $\sim$ 54 km (Figure \ref{fig:fig2}). There are similarities between the two mid-latitudes in the spectral distribution of temperature perturbations, like maximum perturbations occurring in smaller $m$-harmonics or longer vertical wavelengths compared to the high $m$-tail. However, the amplitude of the long vertical wavelength band is greater in the southern midlatitude than the northern, 6 K vs. 3.5 K, peaking at a similar altitude: $\sim$ 70 km vs. $\sim$ 65 km. Overall, these results demonstrate that gravity wave activity could exhibit hemispheric difference on Venus, as often observed in other terrestrial planets. 

Figure \ref{fig:fig3} presents at the two representative midlatitudes the corresponding spectral distributions of the gravity wave absolute horizontal momentum flux ($F_{m}$) and of the absolute gravity wave drag ($a_{h,m}$). The fluxes show an initial exponential increase and maximize around 55 km in both the north and south, increasing with vertical wavelength, with large fluxes at longer $\lambda_z$ spreading over a wider range of altitudes in the south. After the maximum, fluxes show a decrease with altitude. Peak fluxes exceed 1 m$^2$ s$^{-2}$ at large $\lambda_z$, even reach 10 m$^2$ s$^{-2}$ in the southern midlatitude, which are, generally, up to 6 orders of magnitude greater than the fluxes at the shorter $\lambda_z$. The drag is negligible at the high-$m$ tail, but increases with increasing vertical wavelength, exceeding 0.003 m s$^{-2}$, which corresponds to $\sim$ 260 m s$^{-1}$ day$^{-1}$ on Earth, for comparison. The momentum flux and drag remain significant in the higher-$m$/shorter-wavelength harmonics at higher altitudes, in particular in the southern profile. Overall, altitude and spectral behavior of the fluxes resembles the variations of the temperature fluctuations.

 Figure \ref{fig:fig4} shows the total values of the absolute gravity wave momentum flux and of the absolute gravity wave drag, $\sum_m F_{m}$ (top row) and $\sum_m a_{h,m}$ (bottom row),  at the mid-latitudes presented in Figure \ref{fig:fig1} (left:north, right:south). The fluxes show a similar initial exponential increase to that described in Figure \ref{fig:fig3} and then a decrease with altitude above the maximum peak at around 55 km in both hemispheres with 7\flux~ and 10\flux~ in the northern and southern midlatitudes, respectively,  The wave drag is negligible below 50 km at these midlatitudes, but increases above, displaying multiple local peaks, above the altitude where the momentum flux begins to decrease.  The total drag in the southern midlatitude exceeds 0.003 m s$^{-2}$, while it is up to 0.003 m s$^{-2}$ in the northern. In the northern midlatitude the flux and drag both drop above their peaks, while elevated flux and drag values are sustained above their peaks in the southern. 


Gravity waves can exhibit spatiotemporal variations in terrestrial planetary atmospheres, which is also expected on Venus. Figure \ref{fig:fig5} shows the total absolute gravity wave momentum flux for various southern and northern hemisphere profiles, separated into low (0--30\degree), mid (30-60\degree), and high latitude (60--90\degree) bins, and daytime (red) and nighttime (blue). The average flux values are overplotted with thick transparent lines, while the instantaneous ones are represented by thin lines. Across all latitudes and local times, the flux tends to increase at low altitudes and maximizes between 50--60 km before beginning to decrease at higher altitudes. The maximum is more pronounced and attains larger values for the southern hemisphere high latitude bin, noting that there are less observed profiles there. The observed increase could be skewed due to the low number of profiles including 50--60 km considered. Although there are occasional significant differences (e.g., at $\sim $ 55 km in northern hemisphere mid-latitudes, the difference is 4 standard deviations away from 0) between the daytime and nighttime averages (see northern hemisphere low and mid-latitudes), these differences do not follow a clear pattern and become less pronounced when more profiles are included in the analysis (see southern hemisphere low and mid latitudes). While a difference between daytime and nighttime profiles is apparent in the southern hemisphere high latitudes, with nighttime fluxes exceeding daytime fluxes at all observed altitudes, both this result and any northern hemisphere comparison are limited by the small number of profiles available, and more observations would be needed to confirm an increase in flux in the nighttime atmosphere. The maximum average flux is $\sim 60$\flux~ and occurs in the southern hemisphere high latitude nighttime bin. The average flux in the other bins tend to maximize between 5--10\flux. Across all bins, the average flux tends to minimize between 0.5--1\flux. For each latitude bin, the standard deviation is evaluated, reaching an average $\sim$5\flux~ across all profiles.


Figure \ref{fig:fig6} presents the total absolute gravity wave drag and the average total drag for the same profiles as Figure \ref{fig:fig5}, separated into northern and southern hemisphere low, middle, and high latitude bins and daytime and nighttime. Gravity wave drag tends to begin increasing between 50--60 km, at the altitude and above where the momentum flux maximizes, and then experiences multiple peaks at higher altitudes. Wave drag often achieves magnitudes between $0.001$ and $0.002$\drag~or $\sim$86\dragE and $\sim$ 170\dragE, respectively, although there are larger peaks, with average drag maximizing between 0.002 and 0.003\drag, or $\sim$ 170\dragE~and $\sim$260\dragE, at low and middle latitudes, with a couple much larger peaks between 0.012 and 0.02\drag~ or $\sim$1032\dragE~ and $\sim$1720\dragE. Drag maximizes above 0.07\drag~at high latitudes. There is no clear separation or pattern of separation between the daytime and nighttime drag profiles, although there is some indication that the daytime drag is greater than the nighttime one at southern low-latitudes and midlatitudes. For each latitude bin, the standard deviation of drag is evaluated, reaching an average of $\sim 10^{-3}$\drag~ across all profiles.

\section{Discussion}
\label{sec:discussion}

Using neutral temperature data from the Akatsuki spacecraft we have retrieved gravity wave activity in terms of temperature fluctuations, horizontal momentum flux, and wave drag (i.e., wave-induced acceleration) and analyzed their spectral behavior (as a function of vertical wavelength) in the middle atmosphere ($\sim$40--90 km) of Venus. Specifically, 138 profiles were analyzed for temperature fluctuations, spectral amplitudes, absolute horizontal momentum flux, and momentum deposition (drag) produced by gravity waves. Gravity waves are routinely detected in all profiles. Our findings show that gravity wave temperature amplitudes are in the order of $\pm 10$ K with peak amplitudes in the 7.5--15 km vertical wavelength band and significant momentum flux (up to 10\flux) and wave drag (up to 0.003--0.03\drag) are present in all profiles. Here we discuss the altitude variation of gravity wave activity and effects on the mean flow, and how they could be explained by using the theory of gravity wave dissipation in planetary middle atmospheres.

\subsection{Variation of Gravity Waves with Altitude}
\begin{figure}
 \begin{centering}
 \hspace*{-1cm}
    \includegraphics[width=1.15\linewidth, height=\textheight, keepaspectratio]{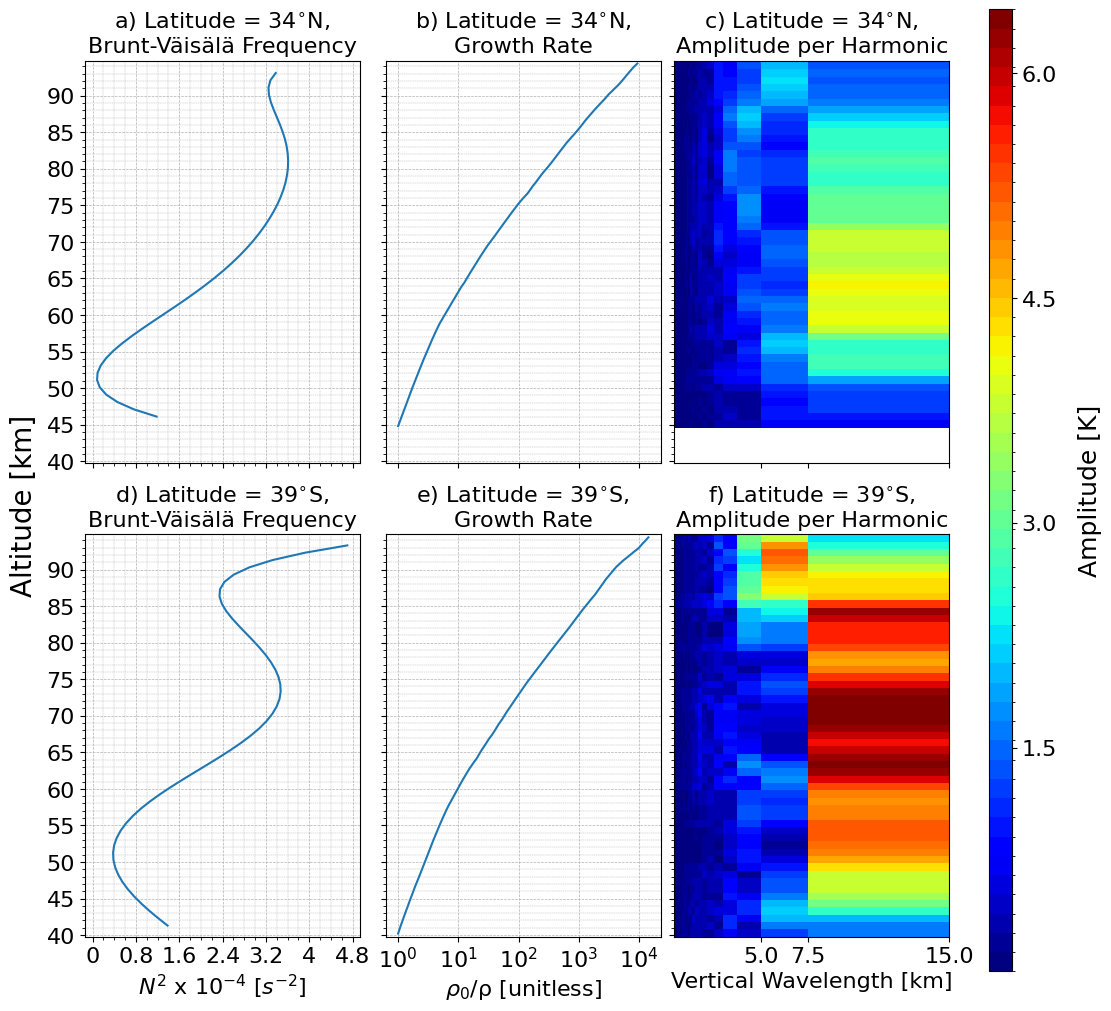}
    \caption{{ (a, c) Derived from Orbit 216, $34^\circ$N profiles. (b, d) Derived from Orbit 118, $39^\circ$S profiles. (a, d) Brunt-Väisälä frequency squared, $N^2$. (b, e) Background density relative to the source density, $\rho_0/\rho(z)$ representing wave growth. (c, f) Amplitude of temperature perturbations, $|T^\prime|$ (K), as a function of altitude and vertical wavelength $\lambda_z$. Vertical wavelength is derived from the wavenumber, $m$. The harmonic of each wave is $m\times20$. Wavelength bins have edges [$\frac{1}{m + 1}$, $\frac{1}{m}$]. The color bar shows the range of temperature perturbation amplitudes, which are plotted at each altitude and harmonic. White spaces indicate altitudes without data coverage.}}
    \label{fig:fig2}
  \end{centering}
\end{figure}

\begin{figure}
\begin{centering}
    \includegraphics[width=1.0\linewidth, height=\textheight, keepaspectratio]{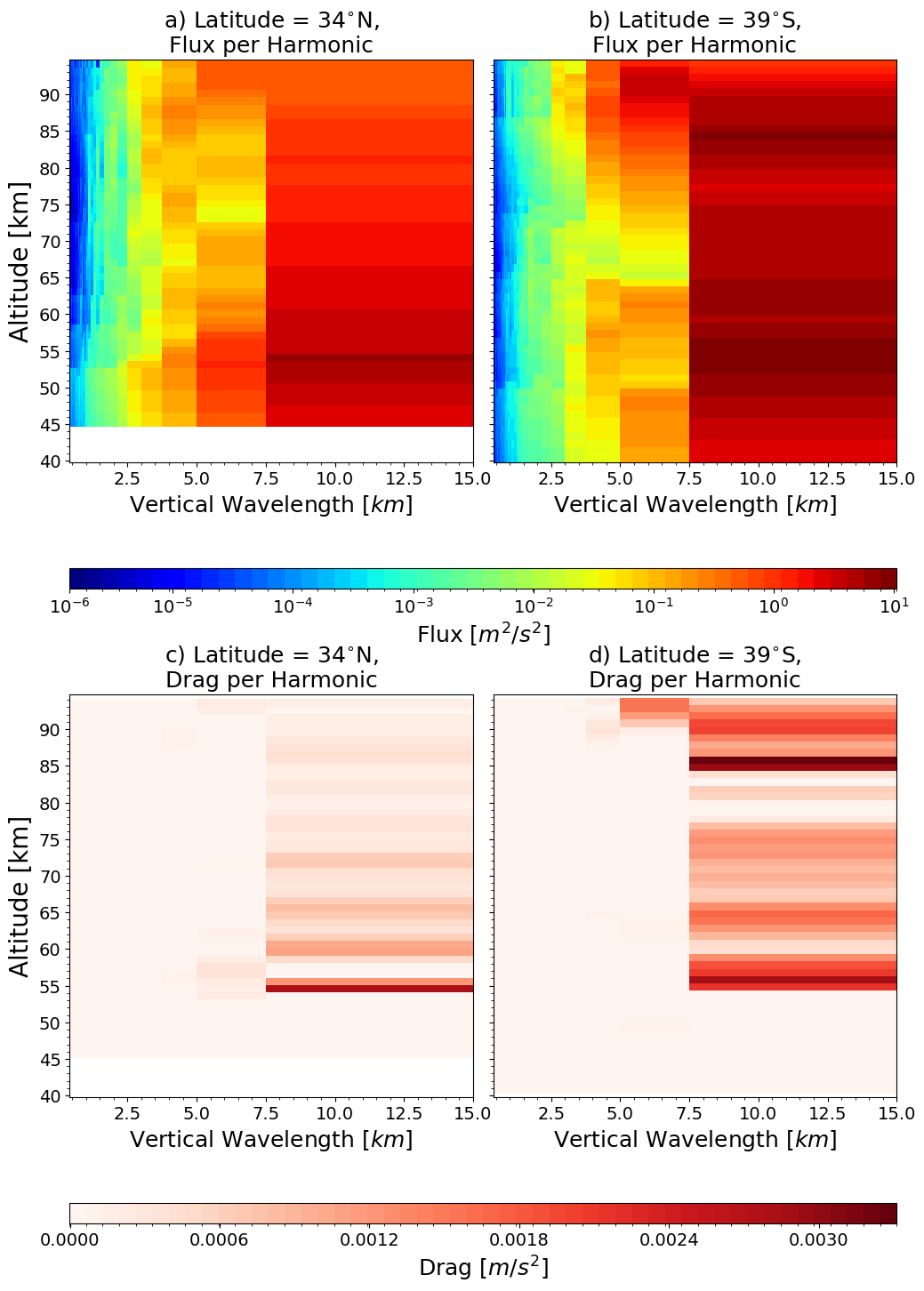}
    \caption{{Gravity wave absolute momentum flux (upper panels a,b) and absolute horizontal gravity wave drag (lower panels c,d) as a function of altitude and vertical wavelength for the northern midlatitude (left panels a and c; Orbit 216, $34^\circ$N) and southern midlatitude (right panels b and d; Orbit 118, $39^\circ$S). Vertical wavelength bin edges are [$\frac{1}{m + 1}$, $\frac{1}{m}$].}}
    \label{fig:fig3}
  \end{centering}
\end{figure}

The Brunt-Väisälä frequency ($N^2$, Figure \ref{fig:fig2}) provides a context for the behavior of upward propagating gravity waves. The minimum in $N^2$, consistent with \citeA{tellmann_small-scale_2012}, implies reduced static stability and a transition toward weak stratification. Such conditions ($N^2 \approx 0$ and $N^2< 0$) suggest convective overturn, which could support convection at the corresponding altitudes as a plausible source of upward-propagating gravity waves.

\begin{figure}
\begin{centering}
    \includegraphics[width=1.0\linewidth, height=\textheight, keepaspectratio]{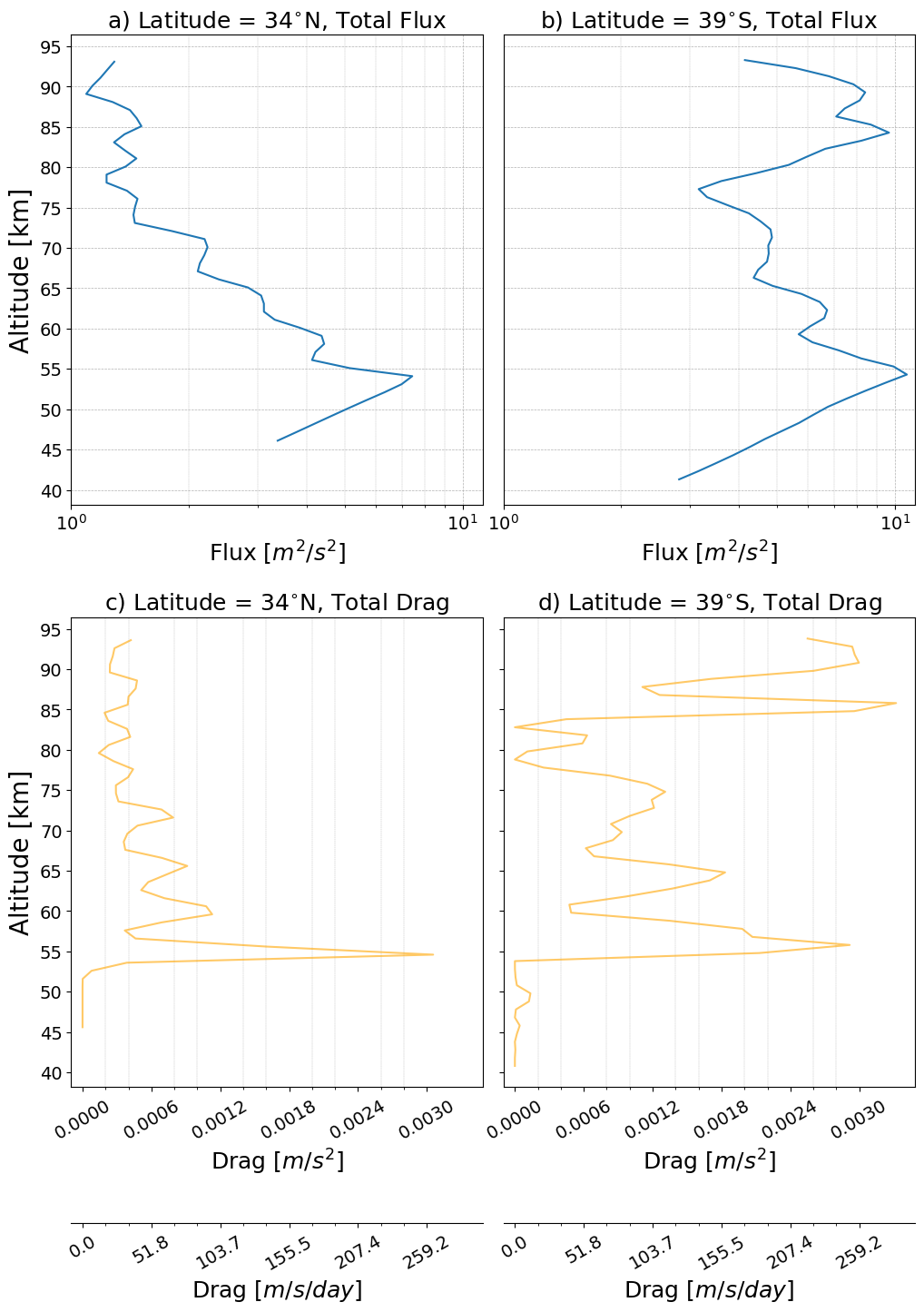}
    \caption{{Altitude variations of the total gravity wave absolute horizontal momentum flux (upper panels) and total absolute horizontal gravity wave drag (lower panels) at the two midlatitudes shown in Figures \ref{fig:fig1}--\ref{fig:fig3}.  There is a secondary $x$-axis for drag in \dragE~ for comparison with Earth, where one day is 24 hours.}}
    \label{fig:fig4}
  \end{centering}
\end{figure}

\begin{figure}
\vspace{-1cm}
\hspace*{-1.8cm}
\begin{centering}
    \includegraphics[width=1.18\linewidth, height=\textheight, keepaspectratio]{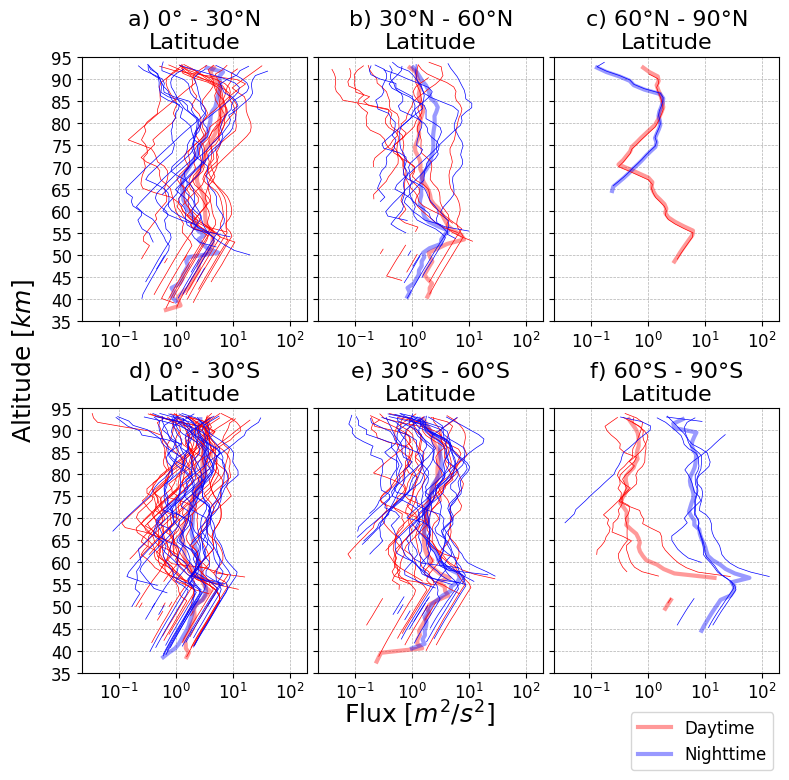}
    \caption{{Total absolute gravity wave horizontal momentum flux is plotted as a function of altitude. Red lines represent daytime profiles ($SZA < 90^\circ$) and blue lines represent nighttime profiles ($SZA > 90^\circ$). Curves are grouped by latitude: (a) Northern hemisphere low latitudes ($0^\circ$--$30^\circ$N), 15 daytime profiles, 11 nighttime profiles; (b) Northern hemisphere middle latitudes ($30^\circ$N--$60^\circ$N), 8 daytime profiles, 8 nighttime profiles; (c) Northern hemisphere high latitudes ($60^\circ$N-- $90^\circ N$), 1 daytime profile, 1 nighttime profile; (d) Southern hemisphere low latitudes ($0^\circ$--$30^\circ$S), 27 daytime profiles, 22 profiles nighttime profiles; (e) Southern hemisphere middle latitudes ($30^\circ$S--$60^\circ$S), 15 daytime profiles, 21 nighttime profiles; (f) Southern hemisphere high latitudes ($60^\circ$S--$90^\circ$S), 4 daytime profile, 5 nighttime profiles. The thick lines are the mean total momentum flux across all daytime or nighttime profiles in a latitude bin. Profiles with data gaps at a certain altitude are excluded from the average at that altitude. The $x$-axis uses a logarithmic scale.}}
    \label{fig:fig5}
    \end{centering}
\end{figure}

\begin{figure}
\vspace{-1cm}
\hspace*{-1.8cm}
\begin{centering}
    \includegraphics[width=1.15\linewidth, height=\textheight, keepaspectratio]{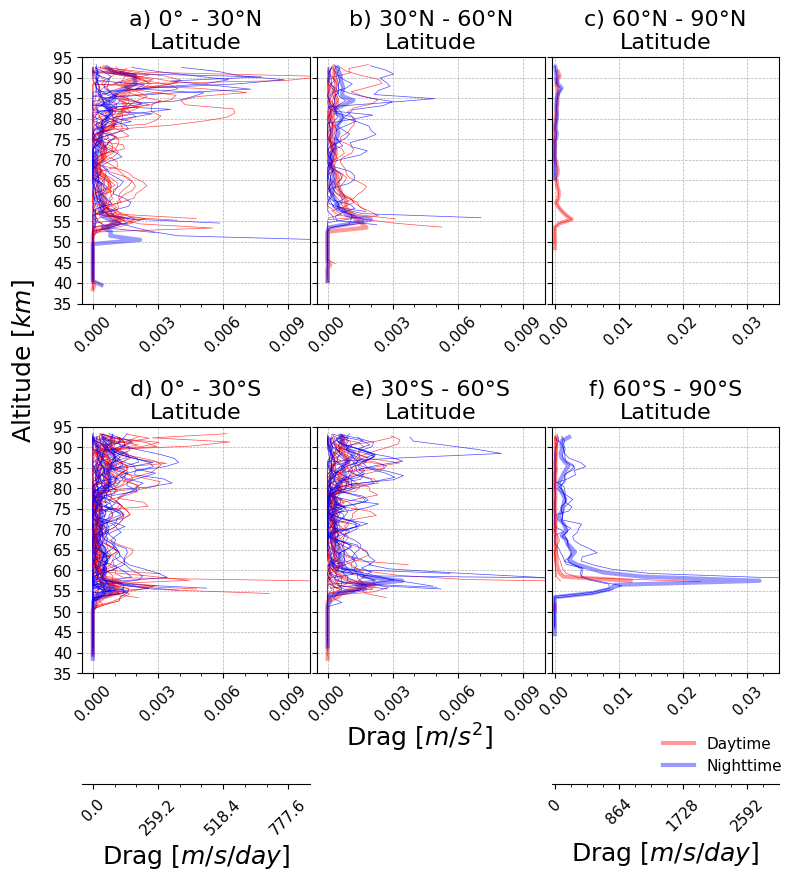}
    \caption{{Total absolute horizontal gravity wave drag in \drag~ is plotted as a function of altitude. Red lines represent daytime profiles ($SZA < 90^\circ$) and blue lines represent nighttime profiles ($SZA>90^\circ$). Curves are grouped by latitude: (a) Northern hemisphere low latitudes ($0^\circ$--$30^\circ$N); (b) Northern hemisphere middle latitudes ($30^\circ$--$60^\circ$N); (c) Northern hemisphere high latitudes ($60^\circ$--$90^\circ$N); (d) Southern hemisphere low latitudes ($0^\circ$--$30^\circ$S); (e) Southern hemisphere middle latitudes ($30^\circ$--$60^\circ$S); (f) Southern hemisphere high latitudes ($60^\circ$--$90^\circ$S). The number of profiles plotted in each altitude band is the same as Figure \ref{fig:fig5}. The thick lines are the mean total absolute drag across all daytime or nighttime profiles in a latitude bin. Profiles with data gaps at a certain altitude are excluded from the average at that altitude.}}
    \label{fig:fig6}
  \end{centering}
\end{figure}

In our study we quantified gravity activity and effects in terms of the wave-induced temperature fluctuations $T^\prime$, the vertical flux of the absolute horizontal momentum flux $F$, and absolute horizontal gravity wave drag $a_h$ as a function of altitude and spectral behavior in terms of vertical wavelength. How can we interpret the observed vertical profiles of gravity waves? Essentially, two major factors shape the altitude distribution of gravity activity in the atmosphere (1) exponential growth of the wave fluxes (or amplitudes) due to the exponential decrease of the background mean density; and (2) wave damping (or dissipation) and critical level wave filtering. For the altitude variation of the vertical flux of horizontal momentum, $\overline{u'w'}_i(z)$ for a given harmonic $i$, this can be stated as \cite<>[Eqn 1]{Yigit.etal2008_ParameterizationEffects}:
\begin{linenomath*}
\begin{equation}
\label{eq: momentum_flux_definition}
     \overline{u'w'}_i(z) = \overline{u'w'}_i(z_0) \frac{\rho(z_0)}{\rho(z)} \tau_i({z}),
\end{equation}
\end{linenomath*}
where $\overline{u'w'}_i(z_0)$ is the flux of the $i$-th harmonic at a reference (or source) level with the background density $\rho(z_0)$ at that level, and $\tau_i(z)$ is the height-dependent transmissivity (analogous to the concept of optical depth) of that harmonic, which accounts for the dissipation of gravity waves. The total flux and dissipation are obtained by superposing the contributions from all harmonics. In \eqref{eq: momentum_flux_definition} the ratio of the reference density to the height-dependent density is called the wave growth factor, which we calculated in Figure \ref{fig:fig2}b,e at two different midlatitudes. The growth factor increases by 4--5 orders of magnitude from 45--95 km within the middle atmosphere, suggesting that the density decrease would lead to an associated wave flux growth of a factor of $10^4$--$10^5$ in the absence of wave dissipation. For this conservative wave propagation the transmissivity is unity, i.e., $\tau_i(z) = 1$, yielding  
\begin{linenomath*}
\begin{equation}
\label{eq: momentum_flux_definition_conservative}
     \overline{u'w'}(z) = \overline{u'w'}(z_0) \frac{\rho(z_0)}{\rho(z)}.
\end{equation}
\end{linenomath*}
At lower altitudes, when dissipation is negligible gravity waves can propagate conservatively, which is indicated by the steep increase of wave flux from 40 to 55 km (Figures \ref{fig:fig4}-\ref{fig:fig5}). However higher in the middle atmosphere above 55 km, wave dissipation can increase, leading to a variable wave transmissivity, which steadily decreases with altitude, i.e., $\tau<1$, as the wave damping intensifies at higher altitudes. The functional form of the wave transmissivity is given in previous works \cite<e.g.,>{Yigit.etal2008_ParameterizationEffects, Yigit.etal2021_EffectsLatitudeDependent}, which we discuss next.  


The initial increase in perturbation amplitudes (Figures \ref{fig:fig1} and \ref{fig:fig2}) and momentum fluxes (Figures \ref{fig:fig3} and \ref{fig:fig4}) suggest initial exponential growth of wave amplitudes without significant dissipation at low altitudes. The initial steep decrease in density (Figure \ref{fig:fig2}) is consistent with density driven amplification of wave amplitudes and momentum flux at low altitudes. However,  as $N^2$ approaches zero the flux estimate becomes sensitive and could produce an artificial enhancement of the momentum flux, contributing to the observed initial increase.  The momentum flux profiles corrected under the conservative or dissipative assumption still exhibit an initial increase  (reduced relative to the uncorrected case), indicating that the 40--60 km growth is likely a genuine feature. Under conservative wave propagation conditions there exists no momentum deposition by gravity waves to the mean flow, as can be seen by the negligible drag below $\sim$55 km (Figure \ref{fig:fig4}).

\subsection{Mechanism of Gravity Wave Damping in the Venusian Middle Atmosphere}
\label{sec:diss_gwdissipation}
Gravity wave temperature fluctuations and horizontal fluxes cannot grow with altitude indefinitely. At some point at higher altitudes due to dissipative processes and/or critical level filtering by the background winds, gravity waves experience increased damping and produce wave drag and/or are absorbed by the background atmosphere. Attenuation of the momentum flux at higher altitudes contrasts with the initial growth and suggests the onset of dissipation. The onset of attenuation is consistent with prior work suggesting saturation between $\sim$ 60--90 km \cite{ando_vertical_2015, mori_gravity_2021, pramitha_vertical_2025}, as amplitude growth is offset by dissipative processes once waves approach the saturation limit. Gravity waves are attenuated due to a combination of various dissipative mechanisms, such as nonlinear diffusion $\beta_{non}$, molecular viscosity and thermal conduction $\beta_{mol}$, ion-neutral friction $\beta_{ion}$, and radiative damping $\beta_{rad}$
\begin{equation}
    \beta_i(z) = \beta_{non} +\beta_{mol} + \beta_{ion} + \beta_{rad},
\end{equation}
where $\beta_i(z)$ is the height-dependent total dissipation a given gravity wave harmonic experiences. Different dissipative processes act generally at different altitudes and under different background atmospheric conditions. Radiative damping has been often considered as a wave dissipation mechanism on Venus \cite{tellmann_small-scale_2012, ando_vertical_2015, mori_gravity_2021}. However, \citeA{imamura_radiative_1995} suggests that radiative damping would not become a significant source of dissipation until above $\sim 120$ km. Radiative damping could still contribute, but it is unlikely to be the sole explanation for momentum flux attenuation and gravity wave drag given that we focus on altitudes well below $\sim 120$ km. Gravity wave dissipation due to ion-neutral friction, molecular diffusion, and nonlinear wave-wave interactions have been virtually unexplored on Venus. Similar to radiative damping, dissipation of gravity waves in the mesosphere due to ion-neutral friction $\beta_{non}$ is negligible in the observed momentum flux variations, given that the Venusian ionosphere extends from $\sim$120--300 km and in the absence of significant ionization in the mesosphere, the ion-neutral friction will be negligible there. Finally, molecular diffusion $\beta_{mol}$ is unlikely to contribute significantly to the attenuation of momentum flux either, given that our focus is on the middle atmosphere and $\beta_{mol}$ starts becoming important only in the thermosphere, which extends from $\sim$120 to 250--350 km. Therefore, this leaves the nonlinear diffusive damping $\beta_{non}$, which is a proxy for the dissipation of gravity waves due to nonlinear interactions between the different gravity wave harmonics, as the primary source of gravity wave dissipation in the mesosphere. Gravity waves posses a relatively broad spectral characteristics, meaning that their spectrum in the lower atmosphere is composed of a wide range of e.g., phases speeds, momentum fluxes, and amplitudes. As gravity wave amplitudes increase with altitude (Figures \ref{fig:fig1}, \ref{fig:fig2}, \ref{fig:fig3}) the individual harmonics start impinging on each other, depending on their relative vertical scales. This is essentially gravity wave nonlinear interactions $\beta_{non}$. Previous studies demonstrated that nonlinear interactions is the primary source of gravity wave dissipation in Earth's \cite{Medvedev.etal1998_RoleAnisotropic, Yigit.etal2009_ModelingEffects, Yigit.etal2014_SimulatedVariability} and Mars's middle atmospheres \cite{Medvedev.etal2011_EstimatesGravity, Yigit.etal2018_InfluenceGravity, Roeten.etal2022_ImpactsGravity}. It is likely that throughout the Venusian mesosphere gravity waves experience similar nonlinear processes, which provides a plausible explanation for the onset of attenuation. Attenuation beginning within a similar altitude range across profiles suggests a common transition region in the Venusian middle atmosphere. 

Within the overall pattern, individual momentum flux profiles show substantial variations and multiple altitude-localized peaks. These peaks can arise because different harmonics reach nonlinear instability thresholds and attenuate at different altitudes, e.g., due to differences in intrinsic horizontal phase speeds, $\hat{c}_i = c_i -\bar{u}$, where $c_i$ is the horizontal phase speed of a given harmonic and $\bar{u}$ is the background mean wind. In fact, the $\hat{c}_i$ is a key parameter that determines the propagation and dissipation of gravity waves in the whole atmosphere system. First, if the intrinsic phase speed of a given harmonic approaches zero ($\hat{c}_i \to 0)$, all wave damping processes generally intensify, since gravity wave dissipation $\beta_i$ is inversely proportional to $\hat{c}_i$ \cite{Yigit.etal2008_ParameterizationEffects}.  Second, if the wave meets its critical level ($c_i = \bar{u})$, then it is absorbed by the background atmosphere. Therefore, the background atmospheric stratification, i.e., the variation of the winds, can significantly modulate wave damping produced by dissipative processes, for example by viscosity or nonlinearity, etc. The wave drag profiles reflect the momentum flux profiles because drag corresponds to how momentum flux changes with altitude and the momentum flux spectra anticipate the structure of the drag profiles. Since we derive drag from the vertical gradient of the absolute momentum flux, it should be interpreted as a drag magnitude (lower limit), not as directional acceleration of the background atmosphere.

The linear saturation theory based on the convective instability threshold is an incomplete representation of gravity wave saturation in terrestrial middle atmospheres. Gravity waves always experience some degree of nonlinear interactions. The relationship between the linear and the nonlinear saturation can be explained using the Froude number defined for monochromatic, $Fr_m$, and broad spectra $Fr_s$:
\begin{equation}
    Fr_m = \Bigg|\frac{u^\prime}{c_i - \bar{u}(z)}\Bigg|, \qquad
    Fr_s = \Bigg|\frac{\sigma_i}{c_i - \bar{u}(z)}\Bigg|,
\end{equation}
where 
\begin{equation}
\sigma_i^2 = \sum_{c_j < c_i} \overline{u_j^{\prime 2}}
\end{equation}
is the wind variations produced by the waves with intrinsic phase speeds smaller than the reference intrinsic phase speed of the harmonic that is nonlinearly damped. Linear wave breaking for a monochromatic harmonic occurs when $Fr_m=1$. However, the same harmonic in the broad spectrum nonlinear case breaks when $Fr_s\rightarrow 1$. Since $\sigma_i$ includes a contribution from at least one (self-interaction) or more harmonics (self-interaction and wave-wave interactions), it is generally larger than the amplitude of a single harmonic,  i.e., $\sigma_i \ge u^\prime_i$. Unlike in the linear saturation case, nonlinear dissipation $\beta_{non}$ is a continuous function that grows when amplitudes of the harmonics grow. Therefore, the overturning of gravity waves is a gradual process and occurs at lower amplitudes for the harmonic that nonlinearly interacts within the spectrum \cite{Yigit.etal2008_ParameterizationEffects, Yigit.Medvedev2013_ExtendingParameterization}.

\subsection{Variability of Gravity Wave Activity}
\label{sec:diss_gw_variability}
We next examine latitude and local-time variations of momentum fluxes and gravity waves drag in connection to potential wave sources and wave damping. The lack of a clear day-night separation in the mesospheric gravity waves horizontal momentum fluxes and drag, especially at low- to middle-latitudes, could indicate that the dominant wave sources and the mesospheric gravity wave dissipation do not exhibit a strong local time dependence, or that local time differences are obscured due to the advection of wave signature by strong zonal winds, causing a lack of correlation with their source regions when wave signatures are detected at higher altitudes \cite{imamura_inverse_2014}. It is also possible that the local time variations of wave sources, gravity wave dissipation, and winds field can balance each other, leading to a weak local time signature in gravity waves.  

On the other hand, the observed Venusian gravity waves exhibit strong latitude-dependence (Figures {\ref{fig:fig5}--\ref{fig:fig6}}). One interpretation of the observed latitudinal dependence of momentum flux and drag is that the associated wave damping is latitude-dependent, especially due to varying background winds. Another explanation is that the higher latitude wave sources are stronger, as suggested in previous studies \cite{tellmann_small-scale_2012, imamura_inverse_2014, ando_vertical_2015, pramitha_vertical_2025, mori_gravity_2021}, leading to larger wave amplitudes and thus larger inferred momentum flux and drag. Larger amplitudes could lead to instability and dissipation at lower altitudes. However, because $N^2$ tends to be lower at middle and high latitudes \cite{pramitha_vertical_2025}, part of the apparent enhancement could reflect increased sensitivity of the flux estimate rather than stronger sources. Overall, the broad latitudinal behavior are consistent with convection as an important source of upward propagating gravity waves, while contributions from sources such as jets or thermal-tide jet-exit regions, especially for shorter vertical wavelengths, cannot be excluded \cite{tellmann_small-scale_2012, sugimoto_generation_2021}.

Taken together, the overarching patterns in the inferred momentum flux and drag magnitudes indicate that gravity waves are a persistent and potentially important contributor to momentum transport in the Venusian middle atmosphere, with magnitudes comparable to those used in models to shape the mean flow  \cite<e.g.,>{sugimoto_generation_2021}.

Since we estimate gravity wave drag from the magnitude of the momentum flux, our derived wave drag does not contain directional information. This limitation arises because temperature retrievals alone do not allow the direction of the momentum flux vector to be determined. We found that the wave drag estimated from the absolute momentum flux is a lower limit on the true gravity wave drag (\ref{ap:drag}). If the meridional component of the momentum flux is neglected, the derived drag approaches an approximation of the true drag rather than a strict lower bound. Observations indicate that most gravity waves in the Venusian cloud tops are predominantly oriented in the zonal direction, suggesting that zonal momentum transport is often dominant \cite{peralta_characterization_2008}. Assuming zonal momentum transport is dominant, the estimated drag is likely a reasonable approximation of the magnitude of the wave drag. However, because waves oriented at oblique angles are also observed, meridional momentum transport cannot be entirely neglected, and the derived drag should be interpreted as a lower bound.

A clear separation in vertical wavelengths does not always exist between gravity waves and large-scale waves belonging to the background. Thus, the vertical scales of disturbances associated with tides, planetary waves, and other motions may potentially overlap with those due to GWs. It is desirable to retain the former in the background, but one still has to set a vertical scale that separates gravity waves from the larger-scale features. For this we assumed $\lambda_z=15$ km as the upper limit for gravity waves. It should be noted that this assumption can potentially overestimate the retrieved net gravity wave activity by including some contributions from larger scale waves, or underestimate gravity wave activity by excluding gravity waves with vertical wavelengths longer than 15 km.

\section{Summary \& Conclusions}
\label{sec:summary}
Using the radio occultation measurements by the Japanese Akatsuki orbiter from 2016--2024, we have retrieved gravity wave-induced temperature fluctuations, absolute gravity wave horizontal momentum flux, and absolute horizontal gravity wave drag, and their local time and latitude variations in the Venusian mesosphere from 40--90 km. The main inferences of this study are the following:
\begin{itemize}
    \item The gravity-wave induced temperature fluctuations, absolute horizontal momentum flux and the resulting gravity wave drag are significant throughout the Venusian middle atmosphere. The temperature fluctuations and the relative temperature fluctuations generally increase from a few K ($\pm 2$--$3\%$) at 40--50 km up to $\pm 10$ K ($\pm$6--7\%).  Momentum flux and wave drag profiles suggest initial conservative growth and then the onset of dissipation above 50--60 km, where peak values reach 10\flux~ and exceed 0.003\drag~(or $\sim 260$\dragE~on Earth for comparison), for flux and drag, respectively, depending on the latitude. The dissipation with altitude is common to most profiles, suggesting that these are overarching features of the Venusian middle atmosphere. 

    \item Spectral analysis of the gravity waves show that the longer vertical wavelength gravity waves (7.5--15 km) dominate the magnitudes of the fluctuations, fluxes, and drag.
    
    \item The latitude dependence of gravity wave momentum flux is conspicuous with strong fluxes and drag at higher latitudes, which could be hotspots in terms of gravity wave generation and propagation. Globally the absolute horizontal gravity wave momentum flux maximizes in the southern high-latitudes at night with more than 50\flux~ and the drag exceeds 0.03\drag~(or 2500\dragE on Earth for comparison) around 55 km.  

    \item Systematic decrease with altitude of momentum flux and wave drag suggests that wave dissipation plays an important role in shaping Venusian mesospheric gravity waves over a wide range of altitudes above 50 km.
    
    \item Radiative damping, molecular diffusion and thermal conduction, and ion-neutral friction are unlikely to contribute appreciably to the dissipation of gravity waves in the Venusian middle atmosphere. On the other hand, nonlinear dissipation, produced by the wave-wave interaction of the different gravity wave harmonics, is likely the primary mechanism of the decrease of gravity wave amplitudes, horizontal fluxes and production of wave drag in the Venusian mesosphere. 
\end{itemize}

Overall, our results demonstrate that the middle atmosphere of Venus is permeated by internal atmospheric gravity waves of various vertical scales, horizontal fluxes, and amplitudes. NASA's DAVINCI (Deep Atmosphere Venus Investigation of Noble gases, Chemistry, and Imaging) \cite{Garvin.etal2022_RevealingMysteries} mission scheduled for launch in 2030 will survey Venus from its cloud top down to the surface. It will include both an orbiter and descent probe. Especially, the Venus Atmospheric Structure Investigation (VASI), one of the descent spheres of DAVINCI, will provide high-cadence measurements of pressure, temperature, and winds, which could provide complementary measurements of gravity wave activity in the Venusian lower and middle atmosphere. 

\begin{appendix}

\section{Uncertainties}
\label{app:uncertainties}
\subsection{Error Propagation}
\label{app:error_propagation}
Quantifying all sources of uncertainty in radio occultation temperature retrievals is challenging due to the multiple potential sources of error, such as thermal noise, phase noise, errors in the position of the spacecraft, and errors resulting from assumptions in the retrieval algorithm \cite{tripathi_quantification_2022}. We use a Monte Carlo simulation to assess how uncertainties in the retrieved temperature profiles could propagate into our calculated momentum fluxes and wave drag.

Previous studies have estimated the statistical, or random, uncertainty in retrieved Akatsuki radio occultation temperature profiles to be on the order of 0.1 K \cite{imamura_initial_2017, nakamura_akatsuki_2016}.  Using this estimate, \citeA{noguchi_radio_2025} added Gaussian random noise with a standard deviation of 0.1 K to a baseline temperature profile in order to quantify the effect of random temperature errors on estimates of $\lambda_h$ and $\lambda_z$. We adopt a similar Monte Carlo framework, but use larger prescribed temperature errors to evaluate the sensitivity of our momentum flux and drag calculations to uncertainties beyond the reported random error alone. Specifically, we run three simulations in which Gaussian random noise with standard deviations of 1 K, 2 K, and 5 K are added to a representative temperature profile. These values are not intended to represent all possible sources of uncertainty but rather provide a sensitivity test for how different effective temperature uncertainties could influence the retrieved momentum flux and drag values.

After selecting the standard deviations for our Monte Carlo simulations, we ran 100 trials per simulation. This number of trials was chosen to provide a reasonable sample size for estimating the distribution of derived quantities.

We add the Gaussian random noise to a representative temperature profile retrieved on 6 December 2019 at $\sim 16$ hours local time, $\sim 39^\circ S$, and $\sim 70^\circ$ longitude, which is the same southern hemisphere profile used in Figure \ref{fig:fig1}.

\begin{figure}
\vspace{-1cm}
\hspace*{-1.8cm}
\begin{centering}
    \includegraphics[width=1.15\linewidth, height=\textheight, keepaspectratio]{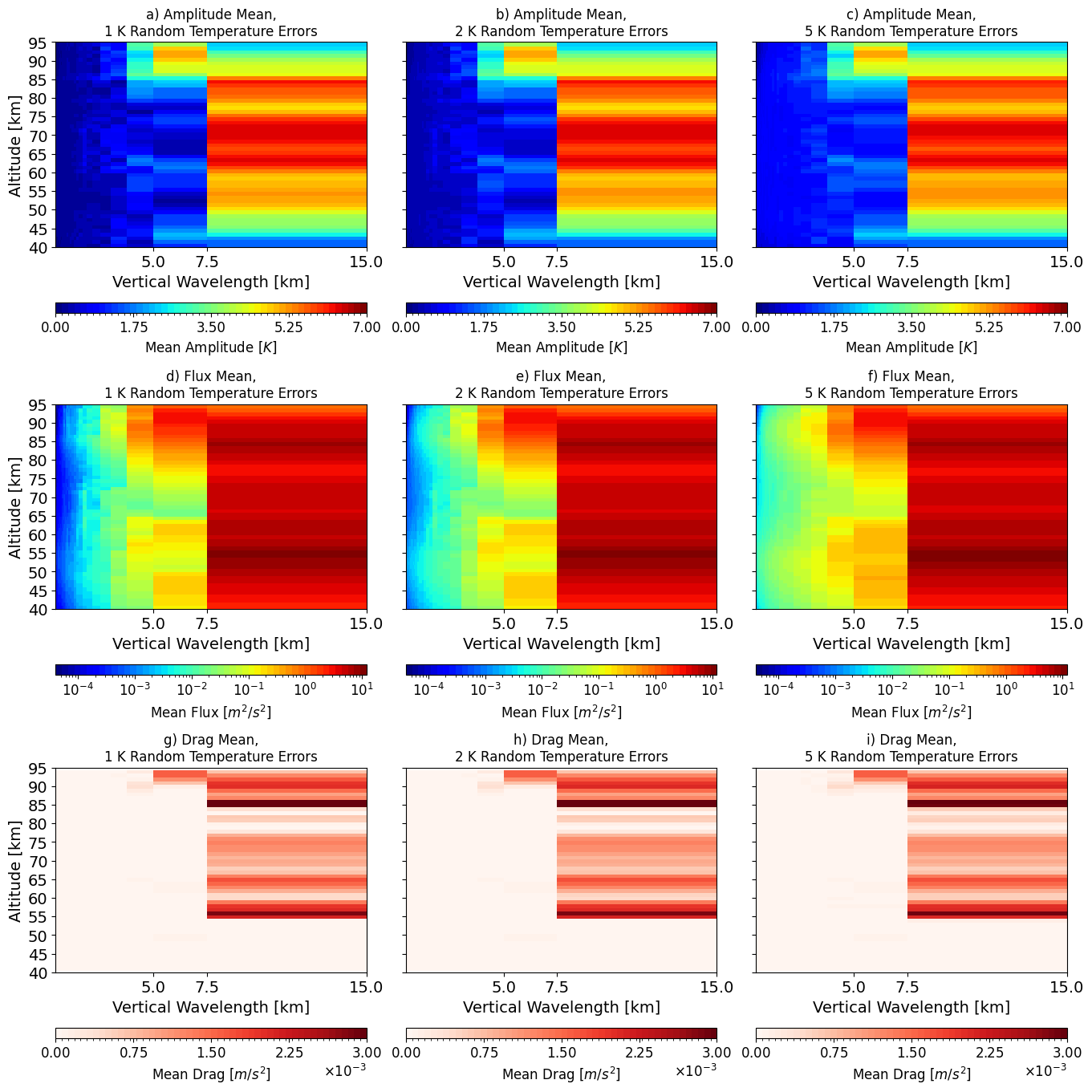}
    \caption{{Mean gravity wave amplitude, absolute momentum flux, and absolute drag as a function of vertical wavelength and altitude, derived from Monte Carlo simulations using the southern hemisphere profile from Figure \ref{fig:fig1}.  Each column corresponds to a different prescribed random temperature error: 1 K (a, d, g), 2 K (b, e, h), and 5 K (c, f, i). Each row shows a different quantity: mean wave amplitude (a–c), mean absolute momentum flux (d–f), and mean absolute wave drag (g–i). }}
    \label{fig:figb1}
  \end{centering}
\end{figure}

To quantify how prescribed temperature errors propagate into uncertainties in the derived wave quantities, Figure \ref{fig:figb1} presents the average wave amplitude, momentum flux, and absolute drag across altitude and vertical wavelength for each of the three error cases. In all cases, neither the mean values nor the region where features like maxima and minima occur differ significantly from the baseline values. For example, for the 5 K temperature error case, the maximum mean momentum flux is $\sim$ 10.6 \flux~ at $\sim$ 54 km, in comparison to the baseline case, where the maximum momentum flux is $\sim 12.2$ \flux~ at $\sim$ 52 km. The consistency between the baseline values and values with added random error demonstrates the robustness of our method.

\begin{figure}
\vspace{-1cm}
\hspace*{-1.8cm}
\begin{centering}
    \includegraphics[width=1.15\linewidth, height=\textheight, keepaspectratio]{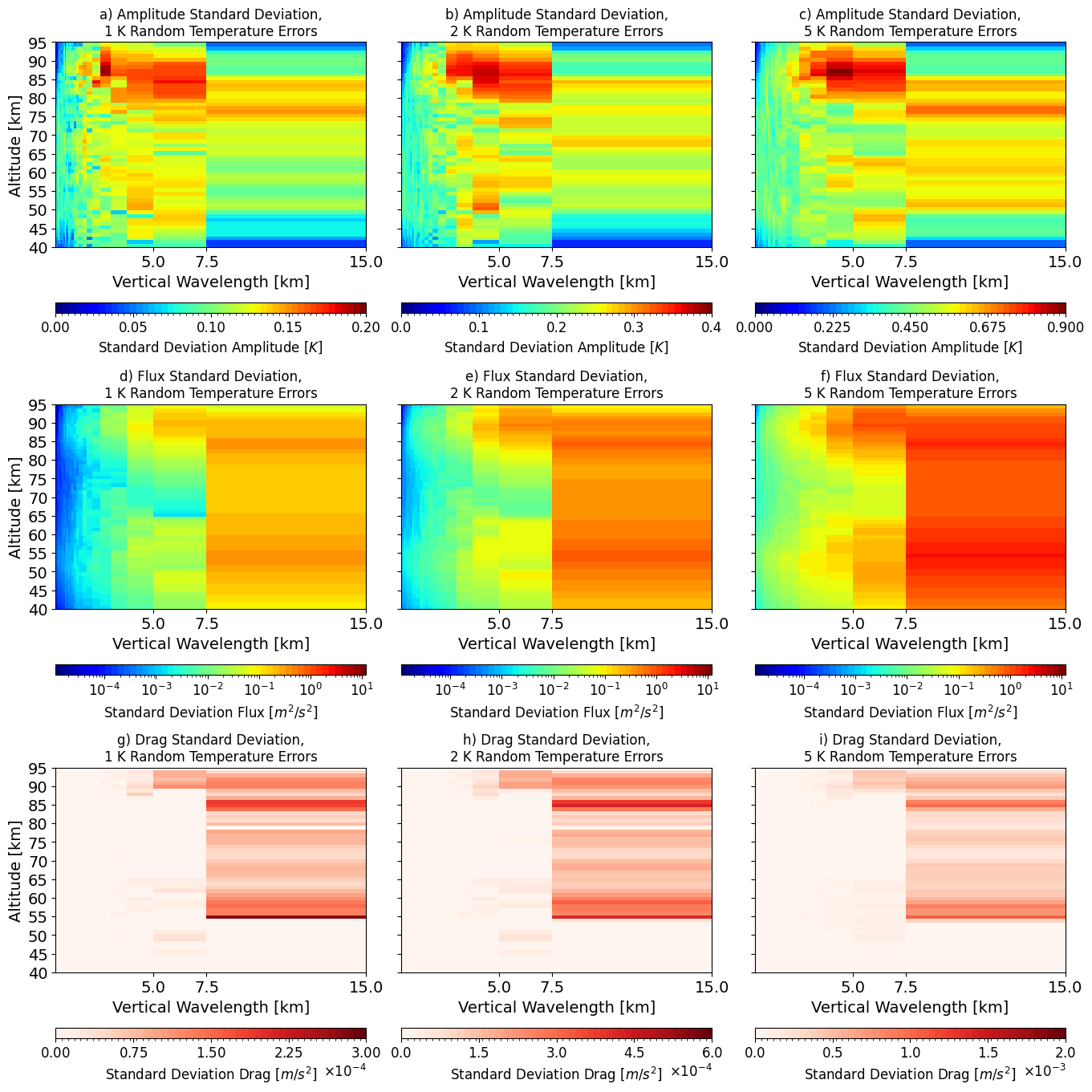}
    \caption{{Standard deviation of the wave amplitude, absolute momentum flux, and absolute drag as a function of vertical wavelength and altitude, derived from Monte Carlo simulations using the southern hemisphere profile from Figure \ref{fig:fig1}. The simulations are the same as those used in Figure \ref{fig:figb1}.  Each column corresponds to a different prescribed random temperature error: 1 K (a, d, g), 2 K (b, e, h), and 5 K (c, f, i). Each row shows a different quantity: standard deviation of wave amplitude (a–c), standard deviation of absolute momentum flux (d–f), and standard deviation of absolute wave drag (g–i).}}
    \label{fig:figb2}
  \end{centering}
\end{figure}

Figure \ref{fig:figb2} presents the standard deviation of wave amplitude, momentum flux, and absolute drag across altitude and vertical wavelength for each of the three error cases. For the 1 K and 2 K temperature errors, the standard deviation of the amplitude, momentum flux, and absolute drag remains $\le$10\% of the mean across most altitude levels. For the 5 K case, this increases to roughly $\sim$ 10--20\% of the mean, reflecting the larger prescribed uncertainty. Although these uncertainties are non-negligible, particularly for the 5 K case, comparison with the mean values in \ref{fig:figb1} demonstrates that they are not a large enough portion of the mean to significantly effect the key features of the derived profiles. For example, the peak momentum flux consistently occurs in the 7.5--15 km vertical wavelength range regardless of the error magnitude.

\begin{figure}
\vspace{-1cm}
\hspace*{-1.8cm}
\begin{centering}
    \includegraphics[width=1.15\linewidth, height=\textheight, keepaspectratio]{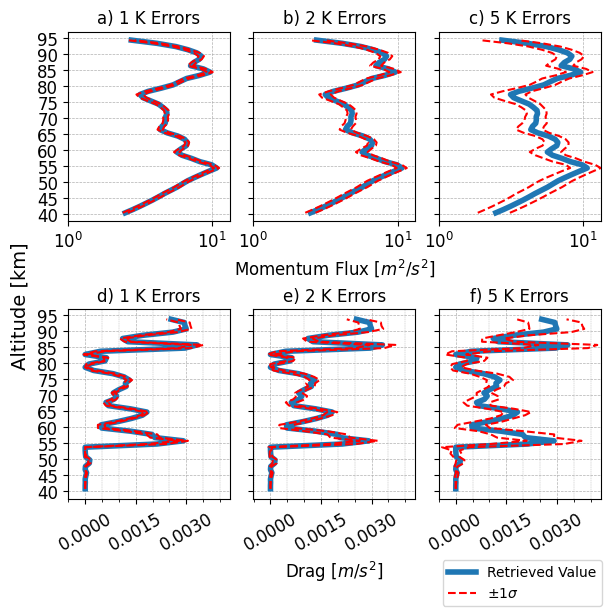}
    \caption{{Retrieved total gravity wave absolute momentum flux (a–c) and absolute drag (d–f) as a function of altitude, with $\pm 1\sigma$ uncertainty bounds derived from Monte Carlo simulations using the southern hemisphere profile from Figure \ref{fig:fig1}. Columns correspond to simulations with 1 K (a, d), 2 K (b, e), and 5 K (c, f) random temperature errors. The solid blue line shows the originally retrieved value and the dashed red lines show the $\pm 1\sigma$ envelope from the Monte Carlo simulations.}}
    \label{fig:figb3}
  \end{centering}
\end{figure}

To better visualize the impact random errors have on the estimated flux and absolute values, we present $\sum F_{k,m} \pm 1\sigma$ and $\sum a_{m}\pm 1\sigma$ for the random errors 1 K, 2 K, and 5 K for the same mid-latitude southern hemisphere profile (Figure \ref{fig:figb3}). As expected, the uncertainties are largest for the 5 K case, but even here the $\pm 1\sigma$ bounds do not shift the flux or drag values enough to change their physical interpretation. In particular, the introduction of random uncertainties does not alter the altitudes at which momentum flux and drag increase or decrease, nor the relative rates of those changes along the profile. Since the vertical structure of flux and drag reflects where and how gravity waves are transferring momentum to the mean flow, the broad preservation of these features across all error magnitudes suggests that random temperature uncertainties of the magnitudes tested are unlikely to fundamentally alter our physical interpretation of gravity wave propagation and momentum deposition, though we note that larger or systematic errors could affect these conclusions.

\subsection{Horizontal Wavelengths}
\label{app:horizontal_wavelength}
The horizontal wavelength of gravity waves in the Venusian atmosphere represents another significant source of uncertainty. Due to the limitations of radio occultation temperature retrievals, the true horizontal wavelength cannot be directly constrained, yet it contributes to the calculated momentum flux and drag values. We therefore present the sensitivity of momentum flux and drag to $\lambda_h$ for waves with $\lambda_z$ between 7.5--15 km.

\begin{figure}
\vspace{-1cm}
\hspace*{-1.8cm}
\begin{centering}
    \includegraphics[width=1.15\linewidth, height=\textheight, keepaspectratio]{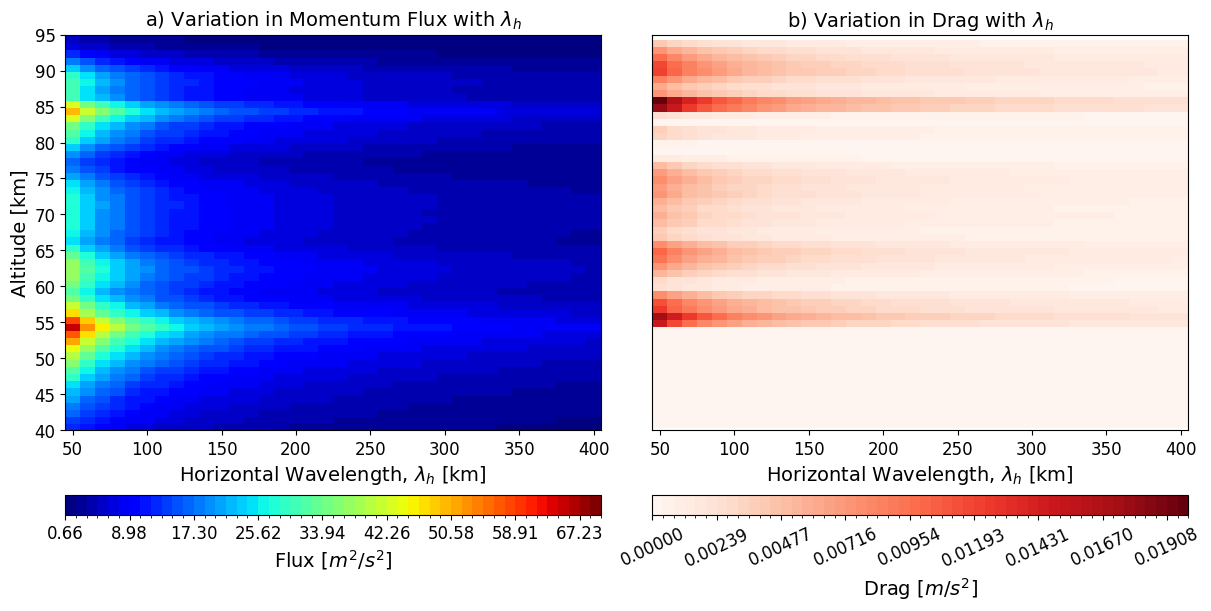}
    \caption{{The variation of the gravity wave absolute momentum flux and absolute drag for the $\lambda_z = $7.5--15 km wavelength bin as a function of horizontal wavelength and altitude. The profile used to derive the momentum flux and drag is the same southern hemisphere profile used in \ref{fig:fig1}. (a) shows the variation of absolute momentum flux with horizontal wavelength. (b) shows the variation of absolute wave drag with horizontal wavelength.}}
    \label{fig:figb4}
  \end{centering}
\end{figure}

As shown in Figure \ref{fig:figb4} and as expressed in Equations \ref{eq:absoluteflux} and \ref{eq:gravity_wave_drag}, momentum flux and drag scale as $\lambda_h^{-1}$, such that waves with larger horizontal wavelengths contribute proportionally less momentum flux and drag than those with shorter horizontal wavelengths.

Observations of Venusian gravity waves indicate a wide range of horizontal wavelengths. Image-based studies of the upper and lower cloud using Venus Express data commonly identify waves with $\lambda_h$ between 50--200 km, although wavelengths outside of this range have been reported, with the maximum reported $\lambda_h$ around 350 km \cite{silva_atmospheric_2024, silva_characterising_2021, peralta_characterization_2008, Piccialli.etal2014_HighLatitudea}. Longer horizontal wavelengths have also been observed, including waves with $\lambda_h$ up to 400 km \cite{Garcia.etal2009_GravityWaves}.

Modeling studies reflect the wide range of possible horizontal wavelengths, with wavelengths between 1--20 km being detected in certain studies \cite{Piccialli.etal2014_HighLatitudea, Lefèvre.etal2017_ThreedimensionalTurbulenceresolving,lefevre_threedimensional_2018} while another study recognized horizontal wavelengths of $\sim 250$ km \cite{sugimoto_generation_2021}.

Figure \ref{fig:figb4} demonstrates that waves with shorter horizontal wavelengths contribute greater momentum flux and drag to the Venusian atmosphere. Given that the majority of observed gravity waves in the Venusian atmosphere have $\lambda_h$ below 300 km \cite{silva_atmospheric_2024, silva_characterising_2021, peralta_characterization_2008, Piccialli.etal2014_HighLatitudea}, the true momentum flux and drag values in the Venusian atmosphere might be greater than those calculated in this study while using a representation horizontal wavelength of 300 km. This further supports that GWs with $\lambda_h$ within the commonly observed range can play a significant role in the circulation of the Venusian atmosphere.

\section{Directional Momentum Deposition}
\label{ap:drag}
The true (directional) gravity wave drag is given by
\begin{equation}
    \mathbf{a}_h = -\frac{1}{\rho(z)}\frac{\partial (\rho \mathbf{F})}{\partial z},
\end{equation}
where $\mathbf{a}_h = (a_x,a_y)$ is composed of the zonal ($a_x$) and meridional  ($a_y$) gravity wave drag components and $\mathbf{F}$ is the vector momentum flux. The wave drag as calculated from the absolute momentum flux we use is given by:
\begin{equation}
    {a_h} = - \frac{1}{\rho(z)}\frac{\partial (\rho |\mathbf{F}|)}{\partial z},
\end{equation}

Using the identity $|\mathbf{F}| = |\sqrt{\mathbf{F} \cdot \mathbf{F}}$, we can rewrite the expression as:
\begin{equation}
    a_h = -\frac{1}{\rho(z)} \bigg( \frac{\mathbf{F}}{|\mathbf{F}|} \cdot \frac{d\mathbf{F}}{dz} \bigg)
\end{equation}
Applying the Cauchy-Schwarz inequality yields, 
\begin{equation}
   \bigg| \frac{\mathbf{F}}{|\mathbf{F}|} \cdot \frac{d\mathbf{F}}{dz}\bigg| 
   \leq \bigg|\frac{\mathbf{F}}{|\mathbf{F}|} \bigg|\cdot\bigg|\frac{d\mathbf{F}}{dz}\bigg|,
\end{equation}
    which implies
 \begin{equation}
     |a_h| \leq |\mathbf{a}_h|.
 \end{equation}
Therefore, the wave drag estimated from the absolute horizontal momentum flux provides a lower bound for the magnitude of the true gravity wave drag in the Venusian atmosphere. 
\end{appendix}
%
%
\begin{acronyms}
\acro{DAVINCI} (Deep Atmosphere Venus Investigation of Noble gases, Chemistry, and Imaging)
\acro{GWs}  Gravity waves
\acro{JAXA} Japan Aerospace Exploration Agency
\acro{RS} Radio Science
\acro{UDSC} Usuda Deep Space Center 
\acro{VASI} Venus Atmospheric Structure Investigation
\acro{VIRTIS} Visible and Infrared Thermal Imaging Spectrometer
\end{acronyms}

\section*{Open Research Section}
The datasets analyzed for this study can be found in the Data ARchives and Transmission System (DARTS), https://data.darts.isas.jaxa.jp/pub/pds3/vco-v-rs-5-occ-v1.0/. The derived momentum fluxes and wave drags used in this study can be found in Zenodo, https://doi.org/10.5281/zenodo.20299533.

\section*{Conflict of Interest declaration}
 The authors declare there are no conflicts of interest for this manuscript.

\acknowledgments
EY was supported by NASA (Grant 80NSSC22K0016) and National Science Foundation (Grant AGS 2330046).

\begin{thebibliography}{}
\bibitem [\protect \citeauthoryear {%
Altieri%
\ \protect \BOthers {.}}{%
Altieri%
\ \protect \BOthers {.}}{%
{\protect \APACyear {2014}}%
}]{%
Altieri.etal2014_ModelingVIRTIS}
\APACinsertmetastar {%
Altieri.etal2014_ModelingVIRTIS}%
\begin{APACrefauthors}%
Altieri, F.%
, Migliorini, A.%
, Zasova, L.%
, Shakun, A.%
, Piccioni, G.%
\BCBL {}\ \BBA {} Bellucci, G.%
\end{APACrefauthors}%
\unskip\
\newblock
\APACrefYearMonthDay{2014}{}{}.
\newblock
{\BBOQ}\APACrefatitle {Modeling {{VIRTIS}}/{{VEX O2}} Nightglow Profiles
  Affected by the Propagation of Gravity Waves in the {{Venus}} Upper
  Mesosphere} {Modeling {{VIRTIS}}/{{VEX O2}} nightglow profiles affected by
  the propagation of gravity waves in the {{Venus}} upper mesosphere}.{\BBCQ}
\newblock
\APACjournalVolNumPages{J. Geophys. Res. Planets}{119}{11}{2300--2316}.
\newblock
\begin{APACrefDOI} \doi{10.1002/2013JE004585} \end{APACrefDOI}
\PrintBackRefs{\CurrentBib}

\bibitem [\protect \citeauthoryear {%
Ando%
\ \protect \BOthers {.}}{%
Ando%
\ \protect \BOthers {.}}{%
{\protect \APACyear {2020}}%
}]{%
Ando.etal2020_ThermalStructure}
\APACinsertmetastar {%
Ando.etal2020_ThermalStructure}%
\begin{APACrefauthors}%
Ando, H.%
, Imamura, T.%
, Tellmann, S.%
, P{\"a}tzold, M.%
, H{\"a}usler, B.%
, Sugimoto, N.%
\BDBL {}Antonita, M.%
\end{APACrefauthors}%
\unskip\
\newblock
\APACrefYearMonthDay{2020}{}{}.
\newblock
{\BBOQ}\APACrefatitle {Thermal Structure of the {{Venusian}} Atmosphere from
  the Sub-Cloud Region to the Mesosphere as Observed by Radio Occultation}
  {Thermal structure of the {{Venusian}} atmosphere from the sub-cloud region
  to the mesosphere as observed by radio occultation}.{\BBCQ}
\newblock
\APACjournalVolNumPages{Sci Rep}{10}{1}{3448}.
\newblock
\begin{APACrefDOI} \doi{10.1038/s41598-020-59278-8} \end{APACrefDOI}
\PrintBackRefs{\CurrentBib}

\bibitem [\protect \citeauthoryear {%
Ando%
, Imamura%
\BCBL {}\ \BBA {} Tsuda%
}{%
Ando%
\ \protect \BOthers {.}}{%
{\protect \APACyear {2012}}%
}]{%
Ando.etal2012_VerticalWavenumber}
\APACinsertmetastar {%
Ando.etal2012_VerticalWavenumber}%
\begin{APACrefauthors}%
Ando, H.%
, Imamura, T.%
\BCBL {}\ \BBA {} Tsuda, T.%
\end{APACrefauthors}%
\unskip\
\newblock
\APACrefYearMonthDay{2012}{}{}.
\newblock
{\BBOQ}\APACrefatitle {Vertical {{Wavenumber Spectra}} of {{Gravity Waves}} in
  the {{Martian Atmosphere Obtained}} from {{Mars Global Surveyor Radio
  Occultation Data}}} {Vertical {{Wavenumber Spectra}} of {{Gravity Waves}} in
  the {{Martian Atmosphere Obtained}} from {{Mars Global Surveyor Radio
  Occultation Data}}}.{\BBCQ}
\newblock
\APACjournalVolNumPages{J. Atmospheric Sci.}{69}{9}{2906--2912}.
\newblock
\begin{APACrefDOI} \doi{10.1175/JAS-D-11-0339.1} \end{APACrefDOI}
\PrintBackRefs{\CurrentBib}

\bibitem [\protect \citeauthoryear {%
Ando%
\ \protect \BOthers {.}}{%
Ando%
\ \protect \BOthers {.}}{%
{\protect \APACyear {2015}}%
}]{%
ando_vertical_2015}
\APACinsertmetastar {%
ando_vertical_2015}%
\begin{APACrefauthors}%
Ando, H.%
, Imamura, T.%
, Tsuda, T.%
, Tellmann, S.%
, Pätzold, M.%
\BCBL {}\ \BBA {} Häusler, B.%
\end{APACrefauthors}%
\unskip\
\newblock
\APACrefYearMonthDay{2015}{{\APACmonth{06}}}{}.
\newblock
{\BBOQ}\APACrefatitle {Vertical {Wavenumber} {Spectra} of {Gravity} {Waves} in
  the {Venus} {Atmosphere} {Obtained} from {Venus} {Express} {Radio}
  {Occultation} {Data}: {Evidence} for {Saturation}} {Vertical {Wavenumber}
  {Spectra} of {Gravity} {Waves} in the {Venus} {Atmosphere} {Obtained} from
  {Venus} {Express} {Radio} {Occultation} {Data}: {Evidence} for
  {Saturation}}.{\BBCQ}
\newblock
\APACjournalVolNumPages{Journal of the Atmospheric
  Sciences}{72}{6}{2318--2329}.
\newblock
\begin{APACrefURL}
  [{2025-11-14}]\url{https://journals.ametsoc.org/doi/10.1175/JAS-D-14-0315.1}
  \end{APACrefURL}
\newblock
\begin{APACrefDOI} \doi{10.1175/JAS-D-14-0315.1} \end{APACrefDOI}
\PrintBackRefs{\CurrentBib}

\bibitem [\protect \citeauthoryear {%
Andrews%
, Holton%
\BCBL {}\ \BBA {} Leovy%
}{%
Andrews%
\ \protect \BOthers {.}}{%
{\protect \APACyear {1987}}%
}]{%
Andrews.etal1987_MiddleAtmosphere}
\APACinsertmetastar {%
Andrews.etal1987_MiddleAtmosphere}%
\begin{APACrefauthors}%
Andrews, D\BPBI G.%
, Holton, J\BPBI R.%
\BCBL {}\ \BBA {} Leovy, C\BPBI B.%
\end{APACrefauthors}%
\unskip\
\newblock
\APACrefYear{1987}.
\newblock
\APACrefbtitle {Middle {{Atmosphere Dynamics}}} {Middle {{Atmosphere
  Dynamics}}}\ (\BVOL~40).
\newblock
\APACaddressPublisher{}{Academic Press}.
\PrintBackRefs{\CurrentBib}

\bibitem [\protect \citeauthoryear {%
Archinal%
\ \protect \BOthers {.}}{%
Archinal%
\ \protect \BOthers {.}}{%
{\protect \APACyear {2018}}%
}]{%
archinal_report_2018}
\APACinsertmetastar {%
archinal_report_2018}%
\begin{APACrefauthors}%
Archinal, B\BPBI A.%
, Acton, C\BPBI H.%
, A’Hearn, M\BPBI F.%
, Conrad, A.%
, Consolmagno, G\BPBI J.%
, Duxbury, T.%
\BDBL {}Williams, I\BPBI P.%
\end{APACrefauthors}%
\unskip\
\newblock
\APACrefYearMonthDay{2018}{{\APACmonth{03}}}{}.
\newblock
{\BBOQ}\APACrefatitle {Report of the {IAU} {Working} {Group} on {Cartographic}
  {Coordinates} and {Rotational} {Elements}: 2015} {Report of the {IAU}
  {Working} {Group} on {Cartographic} {Coordinates} and {Rotational}
  {Elements}: 2015}.{\BBCQ}
\newblock
\APACjournalVolNumPages{Celestial Mechanics and Dynamical
  Astronomy}{130}{3}{22}.
\newblock
\begin{APACrefURL}
  [{2026-05-07}]\url{http://link.springer.com/10.1007/s10569-017-9805-5}
  \end{APACrefURL}
\newblock
\begin{APACrefDOI} \doi{10.1007/s10569-017-9805-5} \end{APACrefDOI}
\PrintBackRefs{\CurrentBib}

\bibitem [\protect \citeauthoryear {%
Baker%
, Schubert%
\BCBL {}\ \BBA {} Jones%
}{%
Baker%
\ \protect \BOthers {.}}{%
{\protect \APACyear {2000}}%
{\protect \APACexlab {{\protect \BCnt {1}}}}}]{%
baker_convectively_2000-1}
\APACinsertmetastar {%
baker_convectively_2000-1}%
\begin{APACrefauthors}%
Baker, R\BPBI D.%
, Schubert, G.%
\BCBL {}\ \BBA {} Jones, P\BPBI W.%
\end{APACrefauthors}%
\unskip\
\newblock
\APACrefYearMonthDay{2000{\protect \BCnt {1}}}{{\APACmonth{01}}}{}.
\newblock
{\BBOQ}\APACrefatitle {Convectively {Generated} {Internal} {Gravity} {Waves} in
  the {Lower} {Atmosphere} of {Venus}. {Part} {II}: {Mean} {Wind} {Shear} and
  {Wave}–{Mean} {Flow} {Interaction}} {Convectively {Generated} {Internal}
  {Gravity} {Waves} in the {Lower} {Atmosphere} of {Venus}. {Part} {II}: {Mean}
  {Wind} {Shear} and {Wave}–{Mean} {Flow} {Interaction}}.{\BBCQ}
\newblock
\APACjournalVolNumPages{Journal of the Atmospheric Sciences}{57}{2}{200--215}.
\newblock
\begin{APACrefURL}
  [{2026-05-07}]\url{http://journals.ametsoc.org/doi/10.1175/1520-0469(2000)057<0200:CGIGWI>2.0.CO;2}
  \end{APACrefURL}
\newblock
\begin{APACrefDOI} \doi{10.1175/1520-0469(2000)057<0200:CGIGWI>2.0.CO;2}
  \end{APACrefDOI}
\PrintBackRefs{\CurrentBib}

\bibitem [\protect \citeauthoryear {%
Baker%
, Schubert%
\BCBL {}\ \BBA {} Jones%
}{%
Baker%
\ \protect \BOthers {.}}{%
{\protect \APACyear {2000}}%
{\protect \APACexlab {{\protect \BCnt {2}}}}}]{%
baker_convectively_2000}
\APACinsertmetastar {%
baker_convectively_2000}%
\begin{APACrefauthors}%
Baker, R\BPBI D.%
, Schubert, G.%
\BCBL {}\ \BBA {} Jones, P\BPBI W.%
\end{APACrefauthors}%
\unskip\
\newblock
\APACrefYearMonthDay{2000{\protect \BCnt {2}}}{{\APACmonth{01}}}{}.
\newblock
{\BBOQ}\APACrefatitle {Convectively {Generated} {Internal} {Gravity} {Waves} in
  the {Lower} {Atmosphere} of {Venus}. {Part} {I}: {No} {Wind} {Shear}}
  {Convectively {Generated} {Internal} {Gravity} {Waves} in the {Lower}
  {Atmosphere} of {Venus}. {Part} {I}: {No} {Wind} {Shear}}.{\BBCQ}
\newblock
\APACjournalVolNumPages{Journal of the Atmospheric Sciences}{57}{2}{184--199}.
\newblock
\begin{APACrefURL}
  [{2026-05-07}]\url{http://journals.ametsoc.org/doi/10.1175/1520-0469(2000)057<0184:CGIGWI>2.0.CO;2}
  \end{APACrefURL}
\newblock
\begin{APACrefDOI} \doi{10.1175/1520-0469(2000)057<0184:CGIGWI>2.0.CO;2}
  \end{APACrefDOI}
\PrintBackRefs{\CurrentBib}

\bibitem [\protect \citeauthoryear {%
Brown%
, Medvedev%
, Starichenko%
, Koskinen%
\BCBL {}\ \BBA {} {M{\"u}ller-Wodarg}%
}{%
Brown%
\ \protect \BOthers {.}}{%
{\protect \APACyear {2022}}%
}]{%
Brown.etal2022_EvidenceGravity}
\APACinsertmetastar {%
Brown.etal2022_EvidenceGravity}%
\begin{APACrefauthors}%
Brown, Z\BPBI L.%
, Medvedev, A\BPBI S.%
, Starichenko, E\BPBI D.%
, Koskinen, T\BPBI T.%
\BCBL {}\ \BBA {} {M{\"u}ller-Wodarg}, I\BPBI C\BPBI F.%
\end{APACrefauthors}%
\unskip\
\newblock
\APACrefYearMonthDay{2022}{}{}.
\newblock
{\BBOQ}\APACrefatitle {Evidence for {{Gravity Waves}} in the {{Thermosphere}}
  of {{Saturn}} and {{Implications}} for {{Global Circulation}}} {Evidence for
  {{Gravity Waves}} in the {{Thermosphere}} of {{Saturn}} and {{Implications}}
  for {{Global Circulation}}}.{\BBCQ}
\newblock
\APACjournalVolNumPages{Geophys. Res. Lett.}{49}{8}{e2021GL097219}.
\newblock
\begin{APACrefDOI} \doi{10.1029/2021GL097219} \end{APACrefDOI}
\PrintBackRefs{\CurrentBib}

\bibitem [\protect \citeauthoryear {%
England%
\ \protect \BOthers {.}}{%
England%
\ \protect \BOthers {.}}{%
{\protect \APACyear {2017}}%
}]{%
England.etal2017_MAVENNGIMS}
\APACinsertmetastar {%
England.etal2017_MAVENNGIMS}%
\begin{APACrefauthors}%
England, S\BPBI L.%
, Liu, G.%
, Yi{\u g}it, E.%
, Mahaffy, P\BPBI R.%
, Elrod, M.%
, Benna, M.%
\BDBL {}Jakosky, B.%
\end{APACrefauthors}%
\unskip\
\newblock
\APACrefYearMonthDay{2017}{}{}.
\newblock
{\BBOQ}\APACrefatitle {{{MAVEN NGIMS}} Observations of Atmospheric Gravity
  Waves in the {{Martian}} Thermosphere} {{{MAVEN NGIMS}} observations of
  atmospheric gravity waves in the {{Martian}} thermosphere}.{\BBCQ}
\newblock
\APACjournalVolNumPages{J. Geophys. Res. Space Physics}{122}{2}{2310--2335}.
\newblock
\begin{APACrefDOI} \doi{10.1002/2016JA023475} \end{APACrefDOI}
\PrintBackRefs{\CurrentBib}

\bibitem [\protect \citeauthoryear {%
Ern%
, Preusse%
, Alexander%
\BCBL {}\ \BBA {} Warner%
}{%
Ern%
\ \protect \BOthers {.}}{%
{\protect \APACyear {2004}}%
}]{%
Ern.etal2004_AbsoluteValues}
\APACinsertmetastar {%
Ern.etal2004_AbsoluteValues}%
\begin{APACrefauthors}%
Ern, M.%
, Preusse, P.%
, Alexander, M\BPBI J.%
\BCBL {}\ \BBA {} Warner, C\BPBI D.%
\end{APACrefauthors}%
\unskip\
\newblock
\APACrefYearMonthDay{2004}{}{}.
\newblock
{\BBOQ}\APACrefatitle {Absolute Values of Gravity Wave Momentum Flux Derived
  from Satellite Data} {Absolute values of gravity wave momentum flux derived
  from satellite data}.{\BBCQ}
\newblock
\APACjournalVolNumPages{J. Geophys. Res.}{109}{D20}{D20103}.
\newblock
\begin{APACrefDOI} \doi{10.1029/2004JD004752} \end{APACrefDOI}
\PrintBackRefs{\CurrentBib}

\bibitem [\protect \citeauthoryear {%
Espadinha%
\ \protect \BOthers {.}}{%
Espadinha%
\ \protect \BOthers {.}}{%
{\protect \APACyear {2026}}%
}]{%
espadinha_wave_2026}
\APACinsertmetastar {%
espadinha_wave_2026}%
\begin{APACrefauthors}%
Espadinha, D.%
, Machado, P.%
, Peralta, J.%
, Imamura, T.%
, Silva, J\BPBI E.%
, Escudero-Coca, P.%
\BCBL {}\ \BBA {} Brasil, F.%
\end{APACrefauthors}%
\unskip\
\newblock
\APACrefYearMonthDay{2026}{{\APACmonth{04}}}{}.
\newblock
{\BBOQ}\APACrefatitle {Wave {Packets} at {Venus}’s {Cloud} {Tops} as {Seen}
  by {Akatsuki}} {Wave {Packets} at {Venus}’s {Cloud} {Tops} as {Seen} by
  {Akatsuki}}.{\BBCQ}
\newblock
\APACjournalVolNumPages{The Planetary Science Journal}{7}{4}{92}.
\newblock
\begin{APACrefURL}
  [{2026-05-07}]\url{https://iopscience.iop.org/article/10.3847/PSJ/ae592e}
  \end{APACrefURL}
\newblock
\begin{APACrefDOI} \doi{10.3847/PSJ/ae592e} \end{APACrefDOI}
\PrintBackRefs{\CurrentBib}

\bibitem [\protect \citeauthoryear {%
Forbes%
, Bruinsma%
, Doornbos%
\BCBL {}\ \BBA {} Zhang%
}{%
Forbes%
\ \protect \BOthers {.}}{%
{\protect \APACyear {2016}}%
}]{%
Forbes.etal2016_GravityWaveinduced}
\APACinsertmetastar {%
Forbes.etal2016_GravityWaveinduced}%
\begin{APACrefauthors}%
Forbes, J\BPBI M.%
, Bruinsma, S\BPBI L.%
, Doornbos, E.%
\BCBL {}\ \BBA {} Zhang, X.%
\end{APACrefauthors}%
\unskip\
\newblock
\APACrefYearMonthDay{2016}{}{}.
\newblock
{\BBOQ}\APACrefatitle {Gravity Wave-Induced Variability of the Middle
  Thermosphere} {Gravity wave-induced variability of the middle
  thermosphere}.{\BBCQ}
\newblock
\APACjournalVolNumPages{J. Geophys. Res. Space Physics}{121}{7}{6914--6923}.
\newblock
\begin{APACrefDOI} \doi{10.1002/2016JA022923} \end{APACrefDOI}
\PrintBackRefs{\CurrentBib}

\bibitem [\protect \citeauthoryear {%
Garcia%
, Drossart%
, Piccioni%
, L{\'o}pez-Valverde%
\BCBL {}\ \BBA {} Occhipinti%
}{%
Garcia%
\ \protect \BOthers {.}}{%
{\protect \APACyear {2009}}%
}]{%
Garcia.etal2009_GravityWaves}
\APACinsertmetastar {%
Garcia.etal2009_GravityWaves}%
\begin{APACrefauthors}%
Garcia, R\BPBI F.%
, Drossart, P.%
, Piccioni, G.%
, L{\'o}pez-Valverde, M.%
\BCBL {}\ \BBA {} Occhipinti, G.%
\end{APACrefauthors}%
\unskip\
\newblock
\APACrefYearMonthDay{2009}{}{}.
\newblock
{\BBOQ}\APACrefatitle {Gravity Waves in the Upper Atmosphere of {{Venus}}
  Revealed by {{CO2}} Nonlocal Thermodynamic Equilibrium Emissions} {Gravity
  waves in the upper atmosphere of {{Venus}} revealed by {{CO2}} nonlocal
  thermodynamic equilibrium emissions}.{\BBCQ}
\newblock
\APACjournalVolNumPages{J. Geophys. Res. Planets}{114}{E5}{}.
\newblock
\begin{APACrefDOI} \doi{10.1029/2008JE003073} \end{APACrefDOI}
\PrintBackRefs{\CurrentBib}

\bibitem [\protect \citeauthoryear {%
Garvin%
\ \protect \BOthers {.}}{%
Garvin%
\ \protect \BOthers {.}}{%
{\protect \APACyear {2022}}%
}]{%
Garvin.etal2022_RevealingMysteries}
\APACinsertmetastar {%
Garvin.etal2022_RevealingMysteries}%
\begin{APACrefauthors}%
Garvin, J\BPBI B.%
, Getty, S\BPBI A.%
, Arney, G\BPBI N.%
, Johnson, N\BPBI M.%
, Kohler, E.%
, Schwer, K\BPBI O.%
\BDBL {}Zolotov, M.%
\end{APACrefauthors}%
\unskip\
\newblock
\APACrefYearMonthDay{2022}{}{}.
\newblock
{\BBOQ}\APACrefatitle {Revealing the {{Mysteries}} of {{Venus}}: {{The DAVINCI
  Mission}}} {Revealing the {{Mysteries}} of {{Venus}}: {{The DAVINCI
  Mission}}}.{\BBCQ}
\newblock
\APACjournalVolNumPages{Planet. Sci. J.}{3}{5}{117}.
\newblock
\begin{APACrefDOI} \doi{10.3847/PSJ/ac63c2} \end{APACrefDOI}
\PrintBackRefs{\CurrentBib}

\bibitem [\protect \citeauthoryear {%
Gavrilov%
\ \BBA {} Kshevetskii%
}{%
Gavrilov%
\ \BBA {} Kshevetskii%
}{%
{\protect \APACyear {2015}}%
}]{%
Gavrilov.Kshevetskii2015_DynamicalThermal}
\APACinsertmetastar {%
Gavrilov.Kshevetskii2015_DynamicalThermal}%
\begin{APACrefauthors}%
Gavrilov, N\BPBI M.%
\BCBT {}\ \BBA {} Kshevetskii, S\BPBI P.%
\end{APACrefauthors}%
\unskip\
\newblock
\APACrefYearMonthDay{2015}{}{}.
\newblock
{\BBOQ}\APACrefatitle {Dynamical and Thermal Effects of Nonsteady Nonlinear
  Acoustic-Gravity Waves Propagating from Tropospheric Sources to the Upper
  Atmosphere} {Dynamical and thermal effects of nonsteady nonlinear
  acoustic-gravity waves propagating from tropospheric sources to the upper
  atmosphere}.{\BBCQ}
\newblock
\APACjournalVolNumPages{Advances in Space Research}{56}{9}{1833--1843}.
\newblock
\begin{APACrefDOI} \doi{10.1016/j.asr.2015.01.033} \end{APACrefDOI}
\PrintBackRefs{\CurrentBib}

\bibitem [\protect \citeauthoryear {%
Heavens%
, Kass%
, Kleinb{\"o}hl%
\BCBL {}\ \BBA {} Schofield%
}{%
Heavens%
\ \protect \BOthers {.}}{%
{\protect \APACyear {2020}}%
}]{%
Heavens.etal2020_MultiannualRecord}
\APACinsertmetastar {%
Heavens.etal2020_MultiannualRecord}%
\begin{APACrefauthors}%
Heavens, N\BPBI G.%
, Kass, D\BPBI M.%
, Kleinb{\"o}hl, A.%
\BCBL {}\ \BBA {} Schofield, J\BPBI T.%
\end{APACrefauthors}%
\unskip\
\newblock
\APACrefYearMonthDay{2020}{}{}.
\newblock
{\BBOQ}\APACrefatitle {A Multiannual Record of Gravity Wave Activity in
  {{Mars}}'s Lower Atmosphere from on-Planet Observations by the {{Mars Climate
  Sounder}}} {A multiannual record of gravity wave activity in {{Mars}}'s lower
  atmosphere from on-planet observations by the {{Mars Climate
  Sounder}}}.{\BBCQ}
\newblock
\APACjournalVolNumPages{Icarus}{341}{}{113630}.
\newblock
\begin{APACrefDOI} \doi{10.1016/j.icarus.2020.113630} \end{APACrefDOI}
\PrintBackRefs{\CurrentBib}

\bibitem [\protect \citeauthoryear {%
Hinson%
\ \BBA {} Jenkins%
}{%
Hinson%
\ \BBA {} Jenkins%
}{%
{\protect \APACyear {1995}}%
}]{%
Hinson.Jenkins1995_MagellanRadio}
\APACinsertmetastar {%
Hinson.Jenkins1995_MagellanRadio}%
\begin{APACrefauthors}%
Hinson, D\BPBI P.%
\BCBT {}\ \BBA {} Jenkins, J\BPBI M.%
\end{APACrefauthors}%
\unskip\
\newblock
\APACrefYearMonthDay{1995}{}{}.
\newblock
{\BBOQ}\APACrefatitle {Magellan {{Radio Occultation Measurements}} of
  {{Atmospheric Waves}} on {{Venus}}} {Magellan {{Radio Occultation
  Measurements}} of {{Atmospheric Waves}} on {{Venus}}}.{\BBCQ}
\newblock
\APACjournalVolNumPages{Icarus}{114}{2}{310--327}.
\newblock
\begin{APACrefDOI} \doi{10.1006/icar.1995.1064} \end{APACrefDOI}
\PrintBackRefs{\CurrentBib}

\bibitem [\protect \citeauthoryear {%
Imamura%
}{%
Imamura%
}{%
{\protect \APACyear {1997}}%
}]{%
imamura_momentum_1997}
\APACinsertmetastar {%
imamura_momentum_1997}%
\begin{APACrefauthors}%
Imamura, T.%
\end{APACrefauthors}%
\unskip\
\newblock
\APACrefYearMonthDay{1997}{{\APACmonth{03}}}{}.
\newblock
{\BBOQ}\APACrefatitle {Momentum balance of the {Venusian} midlatitude
  mesosphere} {Momentum balance of the {Venusian} midlatitude
  mesosphere}.{\BBCQ}
\newblock
\APACjournalVolNumPages{Journal of Geophysical Research:
  Planets}{102}{E3}{6615--6620}.
\newblock
\begin{APACrefURL}
  [{2025-12-03}]\url{https://agupubs.onlinelibrary.wiley.com/doi/10.1029/96JE03882}
  \end{APACrefURL}
\newblock
\begin{APACrefDOI} \doi{10.1029/96JE03882} \end{APACrefDOI}
\PrintBackRefs{\CurrentBib}

\bibitem [\protect \citeauthoryear {%
Imamura%
\ \protect \BOthers {.}}{%
Imamura%
\ \protect \BOthers {.}}{%
{\protect \APACyear {2017}}%
}]{%
imamura_initial_2017}
\APACinsertmetastar {%
imamura_initial_2017}%
\begin{APACrefauthors}%
Imamura, T.%
, Ando, H.%
, Tellmann, S.%
, Pätzold, M.%
, Häusler, B.%
, Yamazaki, A.%
\BDBL {}Nakamura, M.%
\end{APACrefauthors}%
\unskip\
\newblock
\APACrefYearMonthDay{2017}{{\APACmonth{12}}}{}.
\newblock
{\BBOQ}\APACrefatitle {Initial performance of the radio occultation experiment
  in the {Venus} orbiter mission {Akatsuki}} {Initial performance of the radio
  occultation experiment in the {Venus} orbiter mission {Akatsuki}}.{\BBCQ}
\newblock
\APACjournalVolNumPages{Earth, Planets and Space}{69}{1}{137}.
\newblock
\begin{APACrefURL}
  [{2025-10-24}]\url{https://earth-planets-space.springeropen.com/articles/10.1186/s40623-017-0722-3}
  \end{APACrefURL}
\newblock
\begin{APACrefDOI} \doi{10.1186/s40623-017-0722-3} \end{APACrefDOI}
\PrintBackRefs{\CurrentBib}

\bibitem [\protect \citeauthoryear {%
Imamura%
\ \protect \BOthers {.}}{%
Imamura%
\ \protect \BOthers {.}}{%
{\protect \APACyear {2014}}%
}]{%
imamura_inverse_2014}
\APACinsertmetastar {%
imamura_inverse_2014}%
\begin{APACrefauthors}%
Imamura, T.%
, Higuchi, T.%
, Maejima, Y.%
, Takagi, M.%
, Sugimoto, N.%
, Ikeda, K.%
\BCBL {}\ \BBA {} Ando, H.%
\end{APACrefauthors}%
\unskip\
\newblock
\APACrefYearMonthDay{2014}{{\APACmonth{01}}}{}.
\newblock
{\BBOQ}\APACrefatitle {Inverse insolation dependence of {Venus}’ cloud-level
  convection} {Inverse insolation dependence of {Venus}’ cloud-level
  convection}.{\BBCQ}
\newblock
\APACjournalVolNumPages{Icarus}{228}{}{181--188}.
\newblock
\begin{APACrefURL}
  [{2025-11-14}]\url{https://linkinghub.elsevier.com/retrieve/pii/S0019103513004314}
  \end{APACrefURL}
\newblock
\begin{APACrefDOI} \doi{10.1016/j.icarus.2013.10.012} \end{APACrefDOI}
\PrintBackRefs{\CurrentBib}

\bibitem [\protect \citeauthoryear {%
Imamura%
\ \BBA {} Ogawa%
}{%
Imamura%
\ \BBA {} Ogawa%
}{%
{\protect \APACyear {1995}}%
}]{%
imamura_radiative_1995}
\APACinsertmetastar {%
imamura_radiative_1995}%
\begin{APACrefauthors}%
Imamura, T.%
\BCBT {}\ \BBA {} Ogawa, T.%
\end{APACrefauthors}%
\unskip\
\newblock
\APACrefYearMonthDay{1995}{{\APACmonth{02}}}{}.
\newblock
{\BBOQ}\APACrefatitle {Radiative damping of gravity waves in the terrestrial
  planetary atmospheres} {Radiative damping of gravity waves in the terrestrial
  planetary atmospheres}.{\BBCQ}
\newblock
\APACjournalVolNumPages{Geophysical Research Letters}{22}{3}{267--270}.
\newblock
\begin{APACrefURL}
  [{2026-01-07}]\url{https://agupubs.onlinelibrary.wiley.com/doi/10.1029/94GL02998}
  \end{APACrefURL}
\newblock
\begin{APACrefDOI} \doi{10.1029/94GL02998} \end{APACrefDOI}
\PrintBackRefs{\CurrentBib}

\bibitem [\protect \citeauthoryear {%
Imamura%
, Sakurai%
\BCBL {}\ \BBA {} Hinson%
}{%
Imamura%
\ \protect \BOthers {.}}{%
{\protect \APACyear {2024}}%
}]{%
Imamura.etal2024_ShortVertical}
\APACinsertmetastar {%
Imamura.etal2024_ShortVertical}%
\begin{APACrefauthors}%
Imamura, T.%
, Sakurai, R.%
\BCBL {}\ \BBA {} Hinson, D.%
\end{APACrefauthors}%
\unskip\
\newblock
\APACrefYearMonthDay{2024}{}{}.
\newblock
{\BBOQ}\APACrefatitle {Short Vertical Wavelength Gravity Waves in the
  {{Martian}} Polar Lower Atmosphere Observed by Reanalysis of {{Mars Global
  Surveyor}} Radio Occultation Data} {Short vertical wavelength gravity waves
  in the {{Martian}} polar lower atmosphere observed by reanalysis of {{Mars
  Global Surveyor}} radio occultation data}.{\BBCQ}
\newblock
\APACjournalVolNumPages{Icarus}{}{}{116200}.
\newblock
\begin{APACrefDOI} \doi{10.1016/j.icarus.2024.116200} \end{APACrefDOI}
\PrintBackRefs{\CurrentBib}

\bibitem [\protect \citeauthoryear {%
Jesch%
, Medvedev%
, Castellini%
, Yi{\u g}it%
\BCBL {}\ \BBA {} Hartogh%
}{%
Jesch%
\ \protect \BOthers {.}}{%
{\protect \APACyear {2019}}%
}]{%
Jesch.etal2019_DensityFluctuations}
\APACinsertmetastar {%
Jesch.etal2019_DensityFluctuations}%
\begin{APACrefauthors}%
Jesch, D.%
, Medvedev, A\BPBI S.%
, Castellini, F.%
, Yi{\u g}it, E.%
\BCBL {}\ \BBA {} Hartogh, P.%
\end{APACrefauthors}%
\unskip\
\newblock
\APACrefYearMonthDay{2019}{}{}.
\newblock
{\BBOQ}\APACrefatitle {Density {{Fluctuations}} in the {{Lower Thermosphere}}
  of {{Mars Retrieved From}} the {{ExoMars Trace Gas Orbiter}} ({{TGO}})
  {{Aerobraking}}} {Density {{Fluctuations}} in the {{Lower Thermosphere}} of
  {{Mars Retrieved From}} the {{ExoMars Trace Gas Orbiter}} ({{TGO}})
  {{Aerobraking}}}.{\BBCQ}
\newblock
\APACjournalVolNumPages{Atmosphere}{10}{10}{620}.
\newblock
\begin{APACrefDOI} \doi{10.3390/atmos10100620} \end{APACrefDOI}
\PrintBackRefs{\CurrentBib}

\bibitem [\protect \citeauthoryear {%
Kouyama%
\ \protect \BOthers {.}}{%
Kouyama%
\ \protect \BOthers {.}}{%
{\protect \APACyear {2017}}%
}]{%
Kouyama.etal2017_TopographicalLocal}
\APACinsertmetastar {%
Kouyama.etal2017_TopographicalLocal}%
\begin{APACrefauthors}%
Kouyama, T.%
, Imamura, T.%
, Taguchi, M.%
, Fukuhara, T.%
, Sato, T\BPBI M.%
, Yamazaki, A.%
\BDBL {}Nakamura, M.%
\end{APACrefauthors}%
\unskip\
\newblock
\APACrefYearMonthDay{2017}{}{}.
\newblock
{\BBOQ}\APACrefatitle {Topographical and {{Local Time Dependence}} of {{Large
  Stationary Gravity Waves Observed}} at the {{Cloud Top}} of {{Venus}}}
  {Topographical and {{Local Time Dependence}} of {{Large Stationary Gravity
  Waves Observed}} at the {{Cloud Top}} of {{Venus}}}.{\BBCQ}
\newblock
\APACjournalVolNumPages{Geophys. Res. Lett.}{44}{24}{12,098--12,105}.
\newblock
\begin{APACrefDOI} \doi{10.1002/2017GL075792} \end{APACrefDOI}
\PrintBackRefs{\CurrentBib}

\bibitem [\protect \citeauthoryear {%
Leelavathi%
\ \BBA {} Rao%
}{%
Leelavathi%
\ \BBA {} Rao%
}{%
{\protect \APACyear {2024}}%
}]{%
Leelavathi.Rao2024_ComparativeAnalysis}
\APACinsertmetastar {%
Leelavathi.Rao2024_ComparativeAnalysis}%
\begin{APACrefauthors}%
Leelavathi, V.%
\BCBT {}\ \BBA {} Rao, N\BPBI V.%
\end{APACrefauthors}%
\unskip\
\newblock
\APACrefYearMonthDay{2024}{}{}.
\newblock
{\BBOQ}\APACrefatitle {A {{Comparative Analysis}} of {{Gravity Waves}} in
  {{He}} and {{Ar Densities}} in the {{Martian Thermosphere}}} {A {{Comparative
  Analysis}} of {{Gravity Waves}} in {{He}} and {{Ar Densities}} in the
  {{Martian Thermosphere}}}.{\BBCQ}
\newblock
\APACjournalVolNumPages{J. Geophys. Res. Planets}{129}{3}{e2023JE008209}.
\newblock
\begin{APACrefDOI} \doi{10.1029/2023JE008209} \end{APACrefDOI}
\PrintBackRefs{\CurrentBib}

\bibitem [\protect \citeauthoryear {%
Lef{\`e}vre%
, Spiga%
\BCBL {}\ \BBA {} Lebonnois%
}{%
Lef{\`e}vre%
\ \protect \BOthers {.}}{%
{\protect \APACyear {2017}}%
}]{%
Lefèvre.etal2017_ThreedimensionalTurbulenceresolving}
\APACinsertmetastar {%
Lefèvre.etal2017_ThreedimensionalTurbulenceresolving}%
\begin{APACrefauthors}%
Lef{\`e}vre, M.%
, Spiga, A.%
\BCBL {}\ \BBA {} Lebonnois, S.%
\end{APACrefauthors}%
\unskip\
\newblock
\APACrefYearMonthDay{2017}{}{}.
\newblock
{\BBOQ}\APACrefatitle {Three-Dimensional Turbulence-Resolving Modeling of the
  {{Venusian}} Cloud Layer and Induced Gravity Waves} {Three-dimensional
  turbulence-resolving modeling of the {{Venusian}} cloud layer and induced
  gravity waves}.{\BBCQ}
\newblock
\APACjournalVolNumPages{J. Geophys. Res. Planets}{122}{1}{134--149}.
\newblock
\begin{APACrefDOI} \doi{10.1002/2016JE005146} \end{APACrefDOI}
\PrintBackRefs{\CurrentBib}

\bibitem [\protect \citeauthoryear {%
Lefèvre%
, Lebonnois%
\BCBL {}\ \BBA {} Spiga%
}{%
Lefèvre%
\ \protect \BOthers {.}}{%
{\protect \APACyear {2018}}%
}]{%
lefevre_threedimensional_2018}
\APACinsertmetastar {%
lefevre_threedimensional_2018}%
\begin{APACrefauthors}%
Lefèvre, M.%
, Lebonnois, S.%
\BCBL {}\ \BBA {} Spiga, A.%
\end{APACrefauthors}%
\unskip\
\newblock
\APACrefYearMonthDay{2018}{{\APACmonth{10}}}{}.
\newblock
{\BBOQ}\APACrefatitle {Three‐{Dimensional} {Turbulence}‐{Resolving}
  {Modeling} of the {Venusian} {Cloud} {Layer} and {Induced} {Gravity} {Waves}:
  {Inclusion} of {Complete} {Radiative} {Transfer} and {Wind} {Shear}}
  {Three‐{Dimensional} {Turbulence}‐{Resolving} {Modeling} of the
  {Venusian} {Cloud} {Layer} and {Induced} {Gravity} {Waves}: {Inclusion} of
  {Complete} {Radiative} {Transfer} and {Wind} {Shear}}.{\BBCQ}
\newblock
\APACjournalVolNumPages{Journal of Geophysical Research:
  Planets}{123}{10}{2773--2789}.
\newblock
\begin{APACrefURL}
  [{2026-05-01}]\url{https://agupubs.onlinelibrary.wiley.com/doi/10.1029/2018JE005679}
  \end{APACrefURL}
\newblock
\begin{APACrefDOI} \doi{10.1029/2018JE005679} \end{APACrefDOI}
\PrintBackRefs{\CurrentBib}

\bibitem [\protect \citeauthoryear {%
Lefèvre%
, Spiga%
\BCBL {}\ \BBA {} Lebonnois%
}{%
Lefèvre%
\ \protect \BOthers {.}}{%
{\protect \APACyear {2017}}%
}]{%
lefevre_threedimensional_2017}
\APACinsertmetastar {%
lefevre_threedimensional_2017}%
\begin{APACrefauthors}%
Lefèvre, M.%
, Spiga, A.%
\BCBL {}\ \BBA {} Lebonnois, S.%
\end{APACrefauthors}%
\unskip\
\newblock
\APACrefYearMonthDay{2017}{{\APACmonth{01}}}{}.
\newblock
{\BBOQ}\APACrefatitle {Three‐dimensional turbulence‐resolving modeling of
  the {Venusian} cloud layer and induced gravity waves} {Three‐dimensional
  turbulence‐resolving modeling of the {Venusian} cloud layer and induced
  gravity waves}.{\BBCQ}
\newblock
\APACjournalVolNumPages{Journal of Geophysical Research:
  Planets}{122}{1}{134--149}.
\newblock
\begin{APACrefURL}
  [{2026-05-07}]\url{https://agupubs.onlinelibrary.wiley.com/doi/10.1002/2016JE005146}
  \end{APACrefURL}
\newblock
\begin{APACrefDOI} \doi{10.1002/2016JE005146} \end{APACrefDOI}
\PrintBackRefs{\CurrentBib}

\bibitem [\protect \citeauthoryear {%
Leroy%
\ \BBA {} Ingersoll%
}{%
Leroy%
\ \BBA {} Ingersoll%
}{%
{\protect \APACyear {1995}}%
}]{%
leroy_convective_1995}
\APACinsertmetastar {%
leroy_convective_1995}%
\begin{APACrefauthors}%
Leroy, S\BPBI S.%
\BCBT {}\ \BBA {} Ingersoll, A\BPBI P.%
\end{APACrefauthors}%
\unskip\
\newblock
\APACrefYearMonthDay{1995}{{\APACmonth{11}}}{}.
\newblock
{\BBOQ}\APACrefatitle {Convective {Generation} of {Gravity} {Waves} in
  {Venus}'s {Atmosphere}: {Gravity} {Wave} {Spectrum} and {Momentum}
  {Transport}} {Convective {Generation} of {Gravity} {Waves} in {Venus}'s
  {Atmosphere}: {Gravity} {Wave} {Spectrum} and {Momentum} {Transport}}.{\BBCQ}
\newblock
\APACjournalVolNumPages{Journal of the Atmospheric
  Sciences}{52}{21}{3717--3737}.
\newblock
\begin{APACrefURL}
  [{2026-05-07}]\url{http://journals.ametsoc.org/doi/10.1175/1520-0469(1995)052<3717:CGOGWI>2.0.CO;2}
  \end{APACrefURL}
\newblock
\begin{APACrefDOI} \doi{10.1175/1520-0469(1995)052<3717:CGOGWI>2.0.CO;2}
  \end{APACrefDOI}
\PrintBackRefs{\CurrentBib}

\bibitem [\protect \citeauthoryear {%
Leroy%
\ \BBA {} Ingersoll%
}{%
Leroy%
\ \BBA {} Ingersoll%
}{%
{\protect \APACyear {1996}}%
}]{%
leroy_radio_1996}
\APACinsertmetastar {%
leroy_radio_1996}%
\begin{APACrefauthors}%
Leroy, S\BPBI S.%
\BCBT {}\ \BBA {} Ingersoll, A\BPBI P.%
\end{APACrefauthors}%
\unskip\
\newblock
\APACrefYearMonthDay{1996}{{\APACmonth{04}}}{}.
\newblock
{\BBOQ}\APACrefatitle {Radio {Scintillations} in {Venus}'s {Atmosphere}:
  {Application} of a {Theory} of {Gravity} {Wave} {Generation}} {Radio
  {Scintillations} in {Venus}'s {Atmosphere}: {Application} of a {Theory} of
  {Gravity} {Wave} {Generation}}.{\BBCQ}
\newblock
\APACjournalVolNumPages{Journal of the Atmospheric
  Sciences}{53}{7}{1018--1028}.
\newblock
\begin{APACrefURL}
  [{2026-05-07}]\url{http://journals.ametsoc.org/doi/10.1175/1520-0469(1996)053<1018:RSIVAA>2.0.CO;2}
  \end{APACrefURL}
\newblock
\begin{APACrefDOI} \doi{10.1175/1520-0469(1996)053<1018:RSIVAA>2.0.CO;2}
  \end{APACrefDOI}
\PrintBackRefs{\CurrentBib}

\bibitem [\protect \citeauthoryear {%
Liu%
, Millour%
, Forget%
, Gilli%
\BCBL {}\ \protect \BOthers {.}}{%
Liu%
, Millour%
, Forget%
, Gilli%
\BCBL {}\ \protect \BOthers {.}}{%
{\protect \APACyear {2025}}%
}]{%
Liu.etal2025_DiurnalCycle}
\APACinsertmetastar {%
Liu.etal2025_DiurnalCycle}%
\begin{APACrefauthors}%
Liu, J.%
, Millour, E.%
, Forget, F.%
, Gilli, G.%
, Lott, F.%
, Bardet, D.%
\BCBL {}\ \BBA {} Galindo, F\BPBI G.%
\end{APACrefauthors}%
\unskip\
\newblock
\APACrefYearMonthDay{2025}{}{}.
\newblock
{\BBOQ}\APACrefatitle {Diurnal {{Cycle}} of {{Non-Orographic Gravity Waves}}'
  {{Source Altitudes}} and {{Its Impacts}}: {{Tests With Mars Planetary Climate
  Model}}} {Diurnal {{Cycle}} of {{Non-Orographic Gravity Waves}}' {{Source
  Altitudes}} and {{Its Impacts}}: {{Tests With Mars Planetary Climate
  Model}}}.{\BBCQ}
\newblock
\APACjournalVolNumPages{J. Geophys. Res. Planets}{130}{7}{e2024JE008880}.
\newblock
\begin{APACrefDOI} \doi{10.1029/2024JE008880} \end{APACrefDOI}
\PrintBackRefs{\CurrentBib}

\bibitem [\protect \citeauthoryear {%
Liu%
, Millour%
, Forget%
, Lott%
\BCBL {}\ \BBA {} Chaufray%
}{%
Liu%
, Millour%
, Forget%
, Lott%
\BCBL {}\ \BBA {} Chaufray%
}{%
{\protect \APACyear {2025}}%
}]{%
Liu.etal2025_StochasticParameterization}
\APACinsertmetastar {%
Liu.etal2025_StochasticParameterization}%
\begin{APACrefauthors}%
Liu, J.%
, Millour, E.%
, Forget, F.%
, Lott, F.%
\BCBL {}\ \BBA {} Chaufray, J\BHBI Y.%
\end{APACrefauthors}%
\unskip\
\newblock
\APACrefYearMonthDay{2025}{}{}.
\newblock
{\BBOQ}\APACrefatitle {A {{Stochastic Parameterization}} of {{Non-Orographic
  Gravity Waves Induced Mixing}} for {{Mars Planetary Climate Model}}} {A
  {{Stochastic Parameterization}} of {{Non-Orographic Gravity Waves Induced
  Mixing}} for {{Mars Planetary Climate Model}}}.{\BBCQ}
\newblock
\APACjournalVolNumPages{J. Geophys. Res. Planets}{130}{9}{e2025JE009188}.
\newblock
\begin{APACrefDOI} \doi{10.1029/2025JE009188} \end{APACrefDOI}
\PrintBackRefs{\CurrentBib}

\bibitem [\protect \citeauthoryear {%
Medvedev%
\ \BBA {} Klaassen%
}{%
Medvedev%
\ \BBA {} Klaassen%
}{%
{\protect \APACyear {1995}}%
}]{%
Medvedev.Klaassen1995_VerticalEvolution}
\APACinsertmetastar {%
Medvedev.Klaassen1995_VerticalEvolution}%
\begin{APACrefauthors}%
Medvedev, A\BPBI S.%
\BCBT {}\ \BBA {} Klaassen, G\BPBI P.%
\end{APACrefauthors}%
\unskip\
\newblock
\APACrefYearMonthDay{1995}{}{}.
\newblock
{\BBOQ}\APACrefatitle {Vertical Evolution of Gravity Wave Spectra and the
  Parameterization of Associated Wave Drag} {Vertical evolution of gravity wave
  spectra and the parameterization of associated wave drag}.{\BBCQ}
\newblock
\APACjournalVolNumPages{J. Geophys. Res. Atmospheres}{100}{D12}{25841--25853}.
\newblock
\begin{APACrefDOI} \doi{10.1029/95JD02533} \end{APACrefDOI}
\PrintBackRefs{\CurrentBib}

\bibitem [\protect \citeauthoryear {%
Medvedev%
, Klaassen%
\BCBL {}\ \BBA {} Beagley%
}{%
Medvedev%
\ \protect \BOthers {.}}{%
{\protect \APACyear {1998}}%
}]{%
Medvedev.etal1998_RoleAnisotropic}
\APACinsertmetastar {%
Medvedev.etal1998_RoleAnisotropic}%
\begin{APACrefauthors}%
Medvedev, A\BPBI S.%
, Klaassen, G\BPBI P.%
\BCBL {}\ \BBA {} Beagley, S\BPBI R.%
\end{APACrefauthors}%
\unskip\
\newblock
\APACrefYearMonthDay{1998}{}{}.
\newblock
{\BBOQ}\APACrefatitle {On the Role of an Anisotropic Gravity Wave Spectrum in
  Maintaining the Circulation of the Middle Atmosphere} {On the role of an
  anisotropic gravity wave spectrum in maintaining the circulation of the
  middle atmosphere}.{\BBCQ}
\newblock
\APACjournalVolNumPages{Geophys. Res. Lett.}{25}{4}{509--512}.
\newblock
\begin{APACrefDOI} \doi{10.1029/98GL50177} \end{APACrefDOI}
\PrintBackRefs{\CurrentBib}

\bibitem [\protect \citeauthoryear {%
Medvedev%
\ \BBA {} Yi{\u g}it%
}{%
Medvedev%
\ \BBA {} Yi{\u g}it%
}{%
{\protect \APACyear {2019}}%
}]{%
Medvedev.Yigit2019_GravityWaves}
\APACinsertmetastar {%
Medvedev.Yigit2019_GravityWaves}%
\begin{APACrefauthors}%
Medvedev, A\BPBI S.%
\BCBT {}\ \BBA {} Yi{\u g}it, E.%
\end{APACrefauthors}%
\unskip\
\newblock
\APACrefYearMonthDay{2019}{}{}.
\newblock
{\BBOQ}\APACrefatitle {Gravity {{Waves}} in {{Planetary Atmospheres}}: {{Their
  Effects}} and {{Parameterization}} in {{Global Circulation Models}}} {Gravity
  {{Waves}} in {{Planetary Atmospheres}}: {{Their Effects}} and
  {{Parameterization}} in {{Global Circulation Models}}}.{\BBCQ}
\newblock
\APACjournalVolNumPages{Atmosphere}{10}{9}{531}.
\newblock
\begin{APACrefDOI} \doi{10.3390/atmos10090531} \end{APACrefDOI}
\PrintBackRefs{\CurrentBib}

\bibitem [\protect \citeauthoryear {%
Medvedev%
, Yi{\u g}it%
\BCBL {}\ \BBA {} Hartogh%
}{%
Medvedev%
\ \protect \BOthers {.}}{%
{\protect \APACyear {2011}}%
}]{%
Medvedev.etal2011_EstimatesGravity}
\APACinsertmetastar {%
Medvedev.etal2011_EstimatesGravity}%
\begin{APACrefauthors}%
Medvedev, A\BPBI S.%
, Yi{\u g}it, E.%
\BCBL {}\ \BBA {} Hartogh, P.%
\end{APACrefauthors}%
\unskip\
\newblock
\APACrefYearMonthDay{2011}{}{}.
\newblock
{\BBOQ}\APACrefatitle {Estimates of Gravity Wave Drag on {{Mars}}:
  {{Indication}} of a Possible Lower Thermospheric Wind Reversal} {Estimates of
  gravity wave drag on {{Mars}}: {{Indication}} of a possible lower
  thermospheric wind reversal}.{\BBCQ}
\newblock
\APACjournalVolNumPages{Icarus}{211}{1}{909--912}.
\newblock
\begin{APACrefDOI} \doi{10.1016/j.icarus.2010.10.013} \end{APACrefDOI}
\PrintBackRefs{\CurrentBib}

\bibitem [\protect \citeauthoryear {%
Mori%
\ \protect \BOthers {.}}{%
Mori%
\ \protect \BOthers {.}}{%
{\protect \APACyear {2021}}%
}]{%
mori_gravity_2021}
\APACinsertmetastar {%
mori_gravity_2021}%
\begin{APACrefauthors}%
Mori, R.%
, Imamura, T.%
, Ando, H.%
, Häusler, B.%
, Pätzold, M.%
\BCBL {}\ \BBA {} Tellmann, S.%
\end{APACrefauthors}%
\unskip\
\newblock
\APACrefYearMonthDay{2021}{{\APACmonth{09}}}{}.
\newblock
{\BBOQ}\APACrefatitle {Gravity {Wave} {Packets} in the {Venusian} {Atmosphere}
  {Observed} by {Radio} {Occultation} {Experiments}: {Comparison} {With}
  {Saturation} {Theory}} {Gravity {Wave} {Packets} in the {Venusian}
  {Atmosphere} {Observed} by {Radio} {Occultation} {Experiments}: {Comparison}
  {With} {Saturation} {Theory}}.{\BBCQ}
\newblock
\APACjournalVolNumPages{Journal of Geophysical Research:
  Planets}{126}{9}{e2021JE006912}.
\newblock
\begin{APACrefURL}
  [{2025-11-14}]\url{https://agupubs.onlinelibrary.wiley.com/doi/10.1029/2021JE006912}
  \end{APACrefURL}
\newblock
\begin{APACrefDOI} \doi{10.1029/2021JE006912} \end{APACrefDOI}
\PrintBackRefs{\CurrentBib}

\bibitem [\protect \citeauthoryear {%
Murakami%
\ \protect \BOthers {.}}{%
Murakami%
\ \protect \BOthers {.}}{%
{\protect \APACyear {2017}}%
}]{%
murakami_venus_2017}
\APACinsertmetastar {%
murakami_venus_2017}%
\begin{APACrefauthors}%
Murakami, S\BHBI y.%
, Ando, H.%
, Imamura, T.%
, Kleinböhl, A.%
, Suzuki, S.%
, Yamamoto, Y.%
\BCBL {}\ \BBA {} Hashimoto, G\BPBI L.%
\end{APACrefauthors}%
\unskip\
\newblock
\APACrefYearMonthDay{2017}{{\APACmonth{12}}}{}.
\newblock
\APACrefbtitle {Venus {Climate} {Orbiter} {Akatsuki} {RS} {Bending} {Angle} and
  {Temperature}/{Pressure} {Profiles} {PDS3} dataset.} {Venus {Climate}
  {Orbiter} {Akatsuki} {RS} {Bending} {Angle} and {Temperature}/{Pressure}
  {Profiles} {PDS3} dataset.}
\newblock
\APACaddressPublisher{}{Institute of Space and Astronautical Science, Japan
  Aerospace Exploration Agency}.
\newblock
\begin{APACrefURL}
  [{2026-05-07}]\url{https://darts.isas.jaxa.jp/doi/vco/vco-00015.html}
  \end{APACrefURL}
\newblock
\begin{APACrefDOI} \doi{10.17597/ISAS.DARTS/VCO-00015} \end{APACrefDOI}
\PrintBackRefs{\CurrentBib}

\bibitem [\protect \citeauthoryear {%
Nakamura%
\ \protect \BOthers {.}}{%
Nakamura%
\ \protect \BOthers {.}}{%
{\protect \APACyear {2016}}%
}]{%
nakamura_akatsuki_2016}
\APACinsertmetastar {%
nakamura_akatsuki_2016}%
\begin{APACrefauthors}%
Nakamura, M.%
, Imamura, T.%
, Ishii, N.%
, Abe, T.%
, Kawakatsu, Y.%
, Hirose, C.%
\BDBL {}Kamata, Y.%
\end{APACrefauthors}%
\unskip\
\newblock
\APACrefYearMonthDay{2016}{{\APACmonth{12}}}{}.
\newblock
{\BBOQ}\APACrefatitle {{AKATSUKI} returns to {Venus}} {{AKATSUKI} returns to
  {Venus}}.{\BBCQ}
\newblock
\APACjournalVolNumPages{Earth, Planets and Space}{68}{1}{75}.
\newblock
\begin{APACrefURL}
  [{2025-10-24}]\url{http://earth-planets-space.springeropen.com/articles/10.1186/s40623-016-0457-6}
  \end{APACrefURL}
\newblock
\begin{APACrefDOI} \doi{10.1186/s40623-016-0457-6} \end{APACrefDOI}
\PrintBackRefs{\CurrentBib}

\bibitem [\protect \citeauthoryear {%
Nakamura%
\ \protect \BOthers {.}}{%
Nakamura%
\ \protect \BOthers {.}}{%
{\protect \APACyear {2011}}%
}]{%
Nakamura.etal2011_OverviewVenus}
\APACinsertmetastar {%
Nakamura.etal2011_OverviewVenus}%
\begin{APACrefauthors}%
Nakamura, M.%
, Imamura, T.%
, Ishii, N.%
, Abe, T.%
, Satoh, T.%
, Suzuki, M.%
\BDBL {}Hayashi, Y.%
\end{APACrefauthors}%
\unskip\
\newblock
\APACrefYearMonthDay{2011}{}{}.
\newblock
{\BBOQ}\APACrefatitle {Overview of {{Venus}} Orbiter, {{Akatsuki}}} {Overview
  of {{Venus}} orbiter, {{Akatsuki}}}.{\BBCQ}
\newblock
\APACjournalVolNumPages{Earth Planet Sp}{63}{5}{443--457}.
\newblock
\begin{APACrefDOI} \doi{10.5047/eps.2011.02.009} \end{APACrefDOI}
\PrintBackRefs{\CurrentBib}

\bibitem [\protect \citeauthoryear {%
Noguchi%
\ \protect \BOthers {.}}{%
Noguchi%
\ \protect \BOthers {.}}{%
{\protect \APACyear {2025}}%
}]{%
Noguchi.etal2025_RadioScintillation}
\APACinsertmetastar {%
Noguchi.etal2025_RadioScintillation}%
\begin{APACrefauthors}%
Noguchi, K.%
, Hagino, A.%
, Ando, H.%
, Imamura, T.%
, Tellmann, S.%
\BCBL {}\ \BBA {} P{\"a}tzold, M.%
\end{APACrefauthors}%
\unskip\
\newblock
\APACrefYearMonthDay{2025}{}{}.
\newblock
{\BBOQ}\APACrefatitle {Radio {{Scintillation}} and {{Gravity Wave
  Characteristics}} in the {{Venusian Atmosphere}}: {{Insights From Akatsuki
  Radio Occultation}}} {Radio {{Scintillation}} and {{Gravity Wave
  Characteristics}} in the {{Venusian Atmosphere}}: {{Insights From Akatsuki
  Radio Occultation}}}.{\BBCQ}
\newblock
\APACjournalVolNumPages{J. Geophys. Res. Planets}{130}{10}{e2025JE009149}.
\newblock
\begin{APACrefDOI} \doi{10.1029/2025JE009149} \end{APACrefDOI}
\PrintBackRefs{\CurrentBib}

\bibitem [\protect \citeauthoryear {%
Noguchi%
\ \protect \BOthers {.}}{%
Noguchi%
\ \protect \BOthers {.}}{%
{\protect \APACyear {2025}}%
}]{%
noguchi_radio_2025}
\APACinsertmetastar {%
noguchi_radio_2025}%
\begin{APACrefauthors}%
Noguchi, K.%
, Hagino, A.%
, Ando, H.%
, Imamura, T.%
, Tellmann, S.%
\BCBL {}\ \BBA {} Pätzold, M.%
\end{APACrefauthors}%
\unskip\
\newblock
\APACrefYearMonthDay{2025}{{\APACmonth{10}}}{}.
\newblock
{\BBOQ}\APACrefatitle {Radio {Scintillation} and {Gravity} {Wave}
  {Characteristics} in the {Venusian} {Atmosphere}: {Insights} {From}
  {Akatsuki} {Radio} {Occultation}} {Radio {Scintillation} and {Gravity} {Wave}
  {Characteristics} in the {Venusian} {Atmosphere}: {Insights} {From}
  {Akatsuki} {Radio} {Occultation}}.{\BBCQ}
\newblock
\APACjournalVolNumPages{Journal of Geophysical Research:
  Planets}{130}{10}{e2025JE009149}.
\newblock
\begin{APACrefURL}
  [{2025-10-27}]\url{https://agupubs.onlinelibrary.wiley.com/doi/10.1029/2025JE009149}
  \end{APACrefURL}
\newblock
\begin{APACrefDOI} \doi{10.1029/2025JE009149} \end{APACrefDOI}
\PrintBackRefs{\CurrentBib}

\bibitem [\protect \citeauthoryear {%
Nyassor%
\ \protect \BOthers {.}}{%
Nyassor%
\ \protect \BOthers {.}}{%
{\protect \APACyear {2025}}%
}]{%
Nyassor.etal2025_MomentumFlux}
\APACinsertmetastar {%
Nyassor.etal2025_MomentumFlux}%
\begin{APACrefauthors}%
Nyassor, P\BPBI K.%
, Wrasse, C\BPBI M.%
, Paulino, I.%
, Yi{\u g}it, E.%
, {Tsali-Brown}, V\BPBI Y.%
, Buriti, R\BPBI A.%
\BDBL {}Gobbi, D.%
\end{APACrefauthors}%
\unskip\
\newblock
\APACrefYearMonthDay{2025}{}{}.
\newblock
{\BBOQ}\APACrefatitle {Momentum Flux Characteristics of Vertically Propagating
  Gravity Waves} {Momentum flux characteristics of vertically propagating
  gravity waves}.{\BBCQ}
\newblock
\APACjournalVolNumPages{Atmospheric Chem. Phys.}{25}{7}{4053--4082}.
\newblock
\begin{APACrefDOI} \doi{10.5194/acp-25-4053-2025} \end{APACrefDOI}
\PrintBackRefs{\CurrentBib}

\bibitem [\protect \citeauthoryear {%
Peralta%
\ \protect \BOthers {.}}{%
Peralta%
\ \protect \BOthers {.}}{%
{\protect \APACyear {2017}}%
}]{%
Peralta.etal2017_StationaryWaves}
\APACinsertmetastar {%
Peralta.etal2017_StationaryWaves}%
\begin{APACrefauthors}%
Peralta, J.%
, Hueso, R.%
, {S{\'a}nchez-Lavega}, A.%
, Lee, Y\BPBI J.%
, Mu{\~n}oz, A\BPBI G.%
, Kouyama, T.%
\BDBL {}Satoh, T.%
\end{APACrefauthors}%
\unskip\
\newblock
\APACrefYearMonthDay{2017}{}{}.
\newblock
{\BBOQ}\APACrefatitle {Stationary Waves and Slowly Moving Features in the Night
  Upper Clouds of {{Venus}}} {Stationary waves and slowly moving features in
  the night upper clouds of {{Venus}}}.{\BBCQ}
\newblock
\APACjournalVolNumPages{Nat Astron}{1}{8}{0187}.
\newblock
\begin{APACrefDOI} \doi{10.1038/s41550-017-0187} \end{APACrefDOI}
\PrintBackRefs{\CurrentBib}

\bibitem [\protect \citeauthoryear {%
Peralta%
, Hueso%
, {S{\'a}nchez-Lavega}%
\BCBL {}\ \protect \BOthers {.}}{%
Peralta%
, Hueso%
, {S{\'a}nchez-Lavega}%
\BCBL {}\ \protect \BOthers {.}}{%
{\protect \APACyear {2008}}%
}]{%
Peralta.etal2008_CharacterizationMesoscale}
\APACinsertmetastar {%
Peralta.etal2008_CharacterizationMesoscale}%
\begin{APACrefauthors}%
Peralta, J.%
, Hueso, R.%
, {S{\'a}nchez-Lavega}, A.%
, Piccioni, G.%
, Lanciano, O.%
\BCBL {}\ \BBA {} Drossart, P.%
\end{APACrefauthors}%
\unskip\
\newblock
\APACrefYearMonthDay{2008}{}{}.
\newblock
{\BBOQ}\APACrefatitle {Characterization of Mesoscale Gravity Waves in the Upper
  and Lower Clouds of {{Venus}} from {{VEX-VIRTIS}} Images} {Characterization
  of mesoscale gravity waves in the upper and lower clouds of {{Venus}} from
  {{VEX-VIRTIS}} images}.{\BBCQ}
\newblock
\APACjournalVolNumPages{J. Geophys. Res. Planets}{113}{E5}{}.
\newblock
\begin{APACrefDOI} \doi{10.1029/2008JE003185} \end{APACrefDOI}
\PrintBackRefs{\CurrentBib}

\bibitem [\protect \citeauthoryear {%
Peralta%
, Hueso%
, Sánchez‐Lavega%
\BCBL {}\ \protect \BOthers {.}}{%
Peralta%
, Hueso%
, Sánchez‐Lavega%
\BCBL {}\ \protect \BOthers {.}}{%
{\protect \APACyear {2008}}%
}]{%
peralta_characterization_2008}
\APACinsertmetastar {%
peralta_characterization_2008}%
\begin{APACrefauthors}%
Peralta, J.%
, Hueso, R.%
, Sánchez‐Lavega, A.%
, Piccioni, G.%
, Lanciano, O.%
\BCBL {}\ \BBA {} Drossart, P.%
\end{APACrefauthors}%
\unskip\
\newblock
\APACrefYearMonthDay{2008}{{\APACmonth{05}}}{}.
\newblock
{\BBOQ}\APACrefatitle {Characterization of mesoscale gravity waves in the upper
  and lower clouds of {Venus} from {VEX}‐{VIRTIS} images} {Characterization
  of mesoscale gravity waves in the upper and lower clouds of {Venus} from
  {VEX}‐{VIRTIS} images}.{\BBCQ}
\newblock
\APACjournalVolNumPages{Journal of Geophysical Research:
  Planets}{113}{E5}{2008JE003185}.
\newblock
\begin{APACrefURL}
  [{2025-11-14}]\url{https://agupubs.onlinelibrary.wiley.com/doi/10.1029/2008JE003185}
  \end{APACrefURL}
\newblock
\begin{APACrefDOI} \doi{10.1029/2008JE003185} \end{APACrefDOI}
\PrintBackRefs{\CurrentBib}

\bibitem [\protect \citeauthoryear {%
Piccialli%
\ \protect \BOthers {.}}{%
Piccialli%
\ \protect \BOthers {.}}{%
{\protect \APACyear {2014}}%
}]{%
Piccialli.etal2014_HighLatitudea}
\APACinsertmetastar {%
Piccialli.etal2014_HighLatitudea}%
\begin{APACrefauthors}%
Piccialli, A.%
, Titov, D\BPBI V.%
, {Sanchez-Lavega}, A.%
, Peralta, J.%
, Shalygina, O.%
, Markiewicz, W\BPBI J.%
\BCBL {}\ \BBA {} Svedhem, H.%
\end{APACrefauthors}%
\unskip\
\newblock
\APACrefYearMonthDay{2014}{}{}.
\newblock
{\BBOQ}\APACrefatitle {High Latitude Gravity Waves at the {{Venus}} Cloud Tops
  as Observed by the {{Venus Monitoring Camera}} on Board {{Venus Express}}}
  {High latitude gravity waves at the {{Venus}} cloud tops as observed by the
  {{Venus Monitoring Camera}} on board {{Venus Express}}}.{\BBCQ}
\newblock
\APACjournalVolNumPages{Icarus}{227}{}{94--111}.
\newblock
\begin{APACrefDOI} \doi{10.1016/j.icarus.2013.09.012} \end{APACrefDOI}
\PrintBackRefs{\CurrentBib}

\bibitem [\protect \citeauthoryear {%
Pramitha%
\ \BBA {} Mathew%
}{%
Pramitha%
\ \BBA {} Mathew%
}{%
{\protect \APACyear {2025}}%
}]{%
pramitha_vertical_2025}
\APACinsertmetastar {%
pramitha_vertical_2025}%
\begin{APACrefauthors}%
Pramitha, M.%
\BCBT {}\ \BBA {} Mathew, A\BPBI J.%
\end{APACrefauthors}%
\unskip\
\newblock
\APACrefYearMonthDay{2025}{{\APACmonth{05}}}{}.
\newblock
{\BBOQ}\APACrefatitle {Vertical wavenumber spectra of gravity waves in the
  {Venus} atmosphere using \textit{{Akatsuki}} radio occultation profiles and
  comparison with model spectra} {Vertical wavenumber spectra of gravity waves
  in the {Venus} atmosphere using \textit{{Akatsuki}} radio occultation
  profiles and comparison with model spectra}.{\BBCQ}
\newblock
\APACjournalVolNumPages{Monthly Notices of the Royal Astronomical
  Society}{540}{2}{1900--1908}.
\newblock
\begin{APACrefURL}
  [{2025-11-14}]\url{https://academic.oup.com/mnras/article/540/2/1900/8120530}
  \end{APACrefURL}
\newblock
\begin{APACrefDOI} \doi{10.1093/mnras/staf679} \end{APACrefDOI}
\PrintBackRefs{\CurrentBib}

\bibitem [\protect \citeauthoryear {%
Roeten%
\ \protect \BOthers {.}}{%
Roeten%
\ \protect \BOthers {.}}{%
{\protect \APACyear {2022}}%
}]{%
Roeten.etal2022_ImpactsGravity}
\APACinsertmetastar {%
Roeten.etal2022_ImpactsGravity}%
\begin{APACrefauthors}%
Roeten, K\BPBI J.%
, Bougher, S\BPBI W.%
, Yi{\u g}it, E.%
, Medvedev, A\BPBI S.%
, Benna, M.%
\BCBL {}\ \BBA {} Elrod, M\BPBI K.%
\end{APACrefauthors}%
\unskip\
\newblock
\APACrefYearMonthDay{2022}{}{}.
\newblock
{\BBOQ}\APACrefatitle {Impacts of {{Gravity Waves}} in the {{Martian
  Thermosphere}}: {{The Mars Global Ionosphere-Thermosphere Model Coupled
  With}} a {{Whole Atmosphere Gravity Wave Scheme}}} {Impacts of {{Gravity
  Waves}} in the {{Martian Thermosphere}}: {{The Mars Global
  Ionosphere-Thermosphere Model Coupled With}} a {{Whole Atmosphere Gravity
  Wave Scheme}}}.{\BBCQ}
\newblock
\APACjournalVolNumPages{J. Geophys. Res. Planets}{127}{12}{e2022JE007477}.
\newblock
\begin{APACrefDOI} \doi{10.1029/2022JE007477} \end{APACrefDOI}
\PrintBackRefs{\CurrentBib}

\bibitem [\protect \citeauthoryear {%
Seiff%
, Young%
, Haberle%
\BCBL {}\ \BBA {} Houben%
}{%
Seiff%
\ \protect \BOthers {.}}{%
{\protect \APACyear {1992}}%
}]{%
Seiff.etal1992_EvidencesWaves}
\APACinsertmetastar {%
Seiff.etal1992_EvidencesWaves}%
\begin{APACrefauthors}%
Seiff, A.%
, Young, R\BPBI E.%
, Haberle, R.%
\BCBL {}\ \BBA {} Houben, H.%
\end{APACrefauthors}%
\unskip\
\newblock
\APACrefYearMonthDay{1992}{}{}.
\newblock
{\BBOQ}\APACrefatitle {The {{Evidences}} of {{Waves}} in the {{Atmospheres}} of
  {{Venus}} and {{Mars}}} {The {{Evidences}} of {{Waves}} in the
  {{Atmospheres}} of {{Venus}} and {{Mars}}}.{\BBCQ}
\newblock
\BIn{} \APACrefbtitle {Venus and {{Mars}}: {{Atmospheres}}, {{Ionospheres}},
  and {{Solar Wind Interactions}}} {Venus and {{Mars}}: {{Atmospheres}},
  {{Ionospheres}}, and {{Solar Wind Interactions}}}\ (\BPGS\ 73--89).
\newblock
\APACaddressPublisher{}{American Geophysical Union (AGU)}.
\newblock
\begin{APACrefDOI} \doi{10.1029/GM066p0073} \end{APACrefDOI}
\PrintBackRefs{\CurrentBib}

\bibitem [\protect \citeauthoryear {%
Shaposhnikov%
, Medvedev%
, Rodin%
, Yi{\u g}it%
\BCBL {}\ \BBA {} Hartogh%
}{%
Shaposhnikov%
\ \protect \BOthers {.}}{%
{\protect \APACyear {2022}}%
}]{%
Shaposhnikov.etal2022_MartianDust}
\APACinsertmetastar {%
Shaposhnikov.etal2022_MartianDust}%
\begin{APACrefauthors}%
Shaposhnikov, D\BPBI S.%
, Medvedev, A\BPBI S.%
, Rodin, A\BPBI V.%
, Yi{\u g}it, E.%
\BCBL {}\ \BBA {} Hartogh, P.%
\end{APACrefauthors}%
\unskip\
\newblock
\APACrefYearMonthDay{2022}{}{}.
\newblock
{\BBOQ}\APACrefatitle {Martian {{Dust Storms}} and {{Gravity Waves}}:
  {{Disentangling Water Transport}} to the {{Upper Atmosphere}}} {Martian
  {{Dust Storms}} and {{Gravity Waves}}: {{Disentangling Water Transport}} to
  the {{Upper Atmosphere}}}.{\BBCQ}
\newblock
\APACjournalVolNumPages{JGR Planets}{127}{1}{}.
\newblock
\begin{APACrefDOI} \doi{10.1029/2021JE007102} \end{APACrefDOI}
\PrintBackRefs{\CurrentBib}

\bibitem [\protect \citeauthoryear {%
Silva%
\ \protect \BOthers {.}}{%
Silva%
\ \protect \BOthers {.}}{%
{\protect \APACyear {2021}}%
}]{%
Silva.etal2021_CharacterisingAtmospheric}
\APACinsertmetastar {%
Silva.etal2021_CharacterisingAtmospheric}%
\begin{APACrefauthors}%
Silva, J\BPBI E.%
, Machado, P.%
, Peralta, J.%
, Brasil, F.%
, Lebonnois, S.%
\BCBL {}\ \BBA {} Lef{\`e}vre, M.%
\end{APACrefauthors}%
\unskip\
\newblock
\APACrefYearMonthDay{2021}{}{}.
\newblock
{\BBOQ}\APACrefatitle {Characterising Atmospheric Gravity Waves on the
  Nightside Lower Clouds of {{Venus}}: A Systematic Analysis} {Characterising
  atmospheric gravity waves on the nightside lower clouds of {{Venus}}: A
  systematic analysis}.{\BBCQ}
\newblock
\APACjournalVolNumPages{A\&A}{649}{}{A34}.
\newblock
\begin{APACrefDOI} \doi{10.1051/0004-6361/202040193} \end{APACrefDOI}
\PrintBackRefs{\CurrentBib}

\bibitem [\protect \citeauthoryear {%
Silva%
\ \protect \BOthers {.}}{%
Silva%
\ \protect \BOthers {.}}{%
{\protect \APACyear {2021}}%
}]{%
silva_characterising_2021}
\APACinsertmetastar {%
silva_characterising_2021}%
\begin{APACrefauthors}%
Silva, J\BPBI E.%
, Machado, P.%
, Peralta, J.%
, Brasil, F.%
, Lebonnois, S.%
\BCBL {}\ \BBA {} Lefèvre, M.%
\end{APACrefauthors}%
\unskip\
\newblock
\APACrefYearMonthDay{2021}{{\APACmonth{05}}}{}.
\newblock
{\BBOQ}\APACrefatitle {Characterising atmospheric gravity waves on the
  nightside lower clouds of {Venus}: a systematic analysis} {Characterising
  atmospheric gravity waves on the nightside lower clouds of {Venus}: a
  systematic analysis}.{\BBCQ}
\newblock
\APACjournalVolNumPages{Astronomy \& Astrophysics}{649}{}{A34}.
\newblock
\begin{APACrefURL}
  [{2025-10-03}]\url{https://www.aanda.org/10.1051/0004-6361/202040193}
  \end{APACrefURL}
\newblock
\begin{APACrefDOI} \doi{10.1051/0004-6361/202040193} \end{APACrefDOI}
\PrintBackRefs{\CurrentBib}

\bibitem [\protect \citeauthoryear {%
Silva%
\ \protect \BOthers {.}}{%
Silva%
\ \protect \BOthers {.}}{%
{\protect \APACyear {2024}}%
}]{%
silva_atmospheric_2024}
\APACinsertmetastar {%
silva_atmospheric_2024}%
\begin{APACrefauthors}%
Silva, J\BPBI E.%
, Peralta, J.%
, Cardesín-Moinelo, A.%
, Hueso, R.%
, Espadinha, D.%
\BCBL {}\ \BBA {} Lee, Y\BPBI J.%
\end{APACrefauthors}%
\unskip\
\newblock
\APACrefYearMonthDay{2024}{{\APACmonth{06}}}{}.
\newblock
{\BBOQ}\APACrefatitle {Atmospheric gravity waves in {Venus} dayside clouds from
  {VIRTIS}-{M} images} {Atmospheric gravity waves in {Venus} dayside clouds
  from {VIRTIS}-{M} images}.{\BBCQ}
\newblock
\APACjournalVolNumPages{Icarus}{415}{}{116076}.
\newblock
\begin{APACrefURL}
  [{2025-10-03}]\url{https://linkinghub.elsevier.com/retrieve/pii/S0019103524001362}
  \end{APACrefURL}
\newblock
\begin{APACrefDOI} \doi{10.1016/j.icarus.2024.116076} \end{APACrefDOI}
\PrintBackRefs{\CurrentBib}

\bibitem [\protect \citeauthoryear {%
Starichenko%
\ \protect \BOthers {.}}{%
Starichenko%
\ \protect \BOthers {.}}{%
{\protect \APACyear {2021}}%
}]{%
Starichenko.etal2021_GravityWave}
\APACinsertmetastar {%
Starichenko.etal2021_GravityWave}%
\begin{APACrefauthors}%
Starichenko, E\BPBI D.%
, Belyaev, D\BPBI A.%
, Medvedev, A\BPBI S.%
, Fedorova, A\BPBI A.%
, Korablev, O\BPBI I.%
, Trokhimovskiy, A.%
\BDBL {}Hartogh, P.%
\end{APACrefauthors}%
\unskip\
\newblock
\APACrefYearMonthDay{2021}{}{}.
\newblock
{\BBOQ}\APACrefatitle {Gravity {{Wave Activity}} in the {{Martian Atmosphere}}
  at {{Altitudes}} 20--160 Km {{From ACS}}/{{TGO Occultation Measurements}}}
  {Gravity {{Wave Activity}} in the {{Martian Atmosphere}} at {{Altitudes}}
  20--160 km {{From ACS}}/{{TGO Occultation Measurements}}}.{\BBCQ}
\newblock
\APACjournalVolNumPages{J. Geophys. Res. Planets}{126}{8}{e2021JE006899}.
\newblock
\begin{APACrefDOI} \doi{10.1029/2021JE006899} \end{APACrefDOI}
\PrintBackRefs{\CurrentBib}

\bibitem [\protect \citeauthoryear {%
Sugimoto%
\ \protect \BOthers {.}}{%
Sugimoto%
\ \protect \BOthers {.}}{%
{\protect \APACyear {2021}}%
}]{%
sugimoto_generation_2021}
\APACinsertmetastar {%
sugimoto_generation_2021}%
\begin{APACrefauthors}%
Sugimoto, N.%
, Fujisawa, Y.%
, Kashimura, H.%
, Noguchi, K.%
, Kuroda, T.%
, Takagi, M.%
\BCBL {}\ \BBA {} Hayashi, Y\BHBI Y.%
\end{APACrefauthors}%
\unskip\
\newblock
\APACrefYearMonthDay{2021}{{\APACmonth{06}}}{}.
\newblock
{\BBOQ}\APACrefatitle {Generation of gravity waves from thermal tides in the
  {Venus} atmosphere} {Generation of gravity waves from thermal tides in the
  {Venus} atmosphere}.{\BBCQ}
\newblock
\APACjournalVolNumPages{Nature Communications}{12}{1}{3682}.
\newblock
\begin{APACrefURL}
  [{2025-11-14}]\url{https://www.nature.com/articles/s41467-021-24002-1}
  \end{APACrefURL}
\newblock
\begin{APACrefDOI} \doi{10.1038/s41467-021-24002-1} \end{APACrefDOI}
\PrintBackRefs{\CurrentBib}

\bibitem [\protect \citeauthoryear {%
Tellmann%
\ \protect \BOthers {.}}{%
Tellmann%
\ \protect \BOthers {.}}{%
{\protect \APACyear {2012}}%
}]{%
tellmann_small-scale_2012}
\APACinsertmetastar {%
tellmann_small-scale_2012}%
\begin{APACrefauthors}%
Tellmann, S.%
, Häusler, B.%
, Hinson, D.%
, Tyler, G.%
, Andert, T.%
, Bird, M.%
\BDBL {}Remus, S.%
\end{APACrefauthors}%
\unskip\
\newblock
\APACrefYearMonthDay{2012}{{\APACmonth{11}}}{}.
\newblock
{\BBOQ}\APACrefatitle {Small-scale temperature fluctuations seen by the {VeRa}
  {Radio} {Science} {Experiment} on {Venus} {Express}} {Small-scale temperature
  fluctuations seen by the {VeRa} {Radio} {Science} {Experiment} on {Venus}
  {Express}}.{\BBCQ}
\newblock
\APACjournalVolNumPages{Icarus}{221}{2}{471--480}.
\newblock
\begin{APACrefURL}
  [{2025-11-14}]\url{https://linkinghub.elsevier.com/retrieve/pii/S0019103512003429}
  \end{APACrefURL}
\newblock
\begin{APACrefDOI} \doi{10.1016/j.icarus.2012.08.023} \end{APACrefDOI}
\PrintBackRefs{\CurrentBib}

\bibitem [\protect \citeauthoryear {%
Titov%
\ \protect \BOthers {.}}{%
Titov%
\ \protect \BOthers {.}}{%
{\protect \APACyear {2012}}%
}]{%
Titov.etal2012_MorphologyCloud}
\APACinsertmetastar {%
Titov.etal2012_MorphologyCloud}%
\begin{APACrefauthors}%
Titov, D\BPBI V.%
, Markiewicz, W\BPBI J.%
, Ignatiev, N\BPBI I.%
, Song, L.%
, Limaye, S\BPBI S.%
, {Sanchez-Lavega}, A.%
\BDBL {}Moissl, R.%
\end{APACrefauthors}%
\unskip\
\newblock
\APACrefYearMonthDay{2012}{}{}.
\newblock
{\BBOQ}\APACrefatitle {Morphology of the Cloud Tops as Observed by the {{Venus
  Express Monitoring Camera}}} {Morphology of the cloud tops as observed by the
  {{Venus Express Monitoring Camera}}}.{\BBCQ}
\newblock
\APACjournalVolNumPages{Icarus}{217}{2}{682--701}.
\newblock
\begin{APACrefDOI} \doi{10.1016/j.icarus.2011.06.020} \end{APACrefDOI}
\PrintBackRefs{\CurrentBib}

\bibitem [\protect \citeauthoryear {%
Tripathi%
\ \BBA {} Choudhary%
}{%
Tripathi%
\ \BBA {} Choudhary%
}{%
{\protect \APACyear {2022}}%
}]{%
tripathi_quantification_2022}
\APACinsertmetastar {%
tripathi_quantification_2022}%
\begin{APACrefauthors}%
Tripathi, K\BPBI R.%
\BCBT {}\ \BBA {} Choudhary, R\BPBI K.%
\end{APACrefauthors}%
\unskip\
\newblock
\APACrefYearMonthDay{2022}{{\APACmonth{06}}}{}.
\newblock
{\BBOQ}\APACrefatitle {Quantification of {Errors} in the {Planetary}
  {Atmospheric} {Profiles} {Derived} {From} {Radio} {Occultation}
  {Measurements}} {Quantification of {Errors} in the {Planetary} {Atmospheric}
  {Profiles} {Derived} {From} {Radio} {Occultation} {Measurements}}.{\BBCQ}
\newblock
\APACjournalVolNumPages{Earth and Space Science}{9}{6}{e2022EA002326}.
\newblock
\begin{APACrefURL}
  [{2026-03-30}]\url{https://agupubs.onlinelibrary.wiley.com/doi/10.1029/2022EA002326}
  \end{APACrefURL}
\newblock
\begin{APACrefDOI} \doi{10.1029/2022EA002326} \end{APACrefDOI}
\PrintBackRefs{\CurrentBib}

\bibitem [\protect \citeauthoryear {%
Weinstock%
}{%
Weinstock%
}{%
{\protect \APACyear {1975}}%
}]{%
Weinstock1975_NonlinearTheory}
\APACinsertmetastar {%
Weinstock1975_NonlinearTheory}%
\begin{APACrefauthors}%
Weinstock, J.%
\end{APACrefauthors}%
\unskip\
\newblock
\APACrefYearMonthDay{1975}{}{}.
\newblock
{\BBOQ}\APACrefatitle {Nonlinear Theory of Gravity Waves and Enhanced Diffusion
  in the Atmosphere} {Nonlinear theory of gravity waves and enhanced diffusion
  in the atmosphere}.{\BBCQ}
\newblock
\APACjournalVolNumPages{Geophys. Res. Lett.}{2}{10}{453--456}.
\newblock
\begin{APACrefDOI} \doi{10.1029/GL002i010p00453} \end{APACrefDOI}
\PrintBackRefs{\CurrentBib}

\bibitem [\protect \citeauthoryear {%
Weinstock%
}{%
Weinstock%
}{%
{\protect \APACyear {1982}}%
}]{%
Weinstock1982_NonlinearTheory}
\APACinsertmetastar {%
Weinstock1982_NonlinearTheory}%
\begin{APACrefauthors}%
Weinstock, J.%
\end{APACrefauthors}%
\unskip\
\newblock
\APACrefYearMonthDay{1982}{}{}.
\newblock
{\BBOQ}\APACrefatitle {Nonlinear {{Theory}} of {{Gravity Waves}}: {{Momentum
  Deposition}}, {{Generalized Rayleigh Friction}}, and {{Diffusion}}}
  {Nonlinear {{Theory}} of {{Gravity Waves}}: {{Momentum Deposition}},
  {{Generalized Rayleigh Friction}}, and {{Diffusion}}}.{\BBCQ}
\newblock
\APACjournalVolNumPages{J. Atmospheric Sci.}{39}{8}{1698--1710}.
\newblock
\begin{APACrefDOI} \doi{10.1175/1520-0469(1982)039<1698:NTOGWM>2.0.CO;2}
  \end{APACrefDOI}
\PrintBackRefs{\CurrentBib}

\bibitem [\protect \citeauthoryear {%
Yi{\u g}it%
, Aylward%
\BCBL {}\ \BBA {} Medvedev%
}{%
Yi{\u g}it%
\ \protect \BOthers {.}}{%
{\protect \APACyear {2008}}%
}]{%
Yigit.etal2008_ParameterizationEffects}
\APACinsertmetastar {%
Yigit.etal2008_ParameterizationEffects}%
\begin{APACrefauthors}%
Yi{\u g}it, E.%
, Aylward, A\BPBI D.%
\BCBL {}\ \BBA {} Medvedev, A\BPBI S.%
\end{APACrefauthors}%
\unskip\
\newblock
\APACrefYearMonthDay{2008}{}{}.
\newblock
{\BBOQ}\APACrefatitle {Parameterization of the Effects of Vertically
  Propagating Gravity Waves for Thermosphere General Circulation Models:
  {{Sensitivity}} Study} {Parameterization of the effects of vertically
  propagating gravity waves for thermosphere general circulation models:
  {{Sensitivity}} study}.{\BBCQ}
\newblock
\APACjournalVolNumPages{J. Geophys. Res.}{113}{D19}{D19106}.
\newblock
\begin{APACrefDOI} \doi{10.1029/2008JD010135} \end{APACrefDOI}
\PrintBackRefs{\CurrentBib}

\bibitem [\protect \citeauthoryear {%
Yi{\u g}it%
\ \protect \BOthers {.}}{%
Yi{\u g}it%
\ \protect \BOthers {.}}{%
{\protect \APACyear {2015}}%
}]{%
Yigit.etal2015_HighaltitudeGravity}
\APACinsertmetastar {%
Yigit.etal2015_HighaltitudeGravity}%
\begin{APACrefauthors}%
Yi{\u g}it, E.%
, England, S\BPBI L.%
, Liu, G.%
, Medvedev, A\BPBI S.%
, Mahaffy, P\BPBI R.%
, Kuroda, T.%
\BCBL {}\ \BBA {} Jakosky, B\BPBI M.%
\end{APACrefauthors}%
\unskip\
\newblock
\APACrefYearMonthDay{2015}{}{}.
\newblock
{\BBOQ}\APACrefatitle {High-Altitude Gravity Waves in the {{Martian}}
  Thermosphere Observed by {{MAVEN}}/{{NGIMS}} and Modeled by a Gravity Wave
  Scheme} {High-altitude gravity waves in the {{Martian}} thermosphere observed
  by {{MAVEN}}/{{NGIMS}} and modeled by a gravity wave scheme}.{\BBCQ}
\newblock
\APACjournalVolNumPages{Geophys. Res. Lett.}{42}{21}{8993--9000}.
\newblock
\begin{APACrefDOI} \doi{10.1002/2015GL065307} \end{APACrefDOI}
\PrintBackRefs{\CurrentBib}

\bibitem [\protect \citeauthoryear {%
Yi{\u g}it%
\ \BBA {} Medvedev%
}{%
Yi{\u g}it%
\ \BBA {} Medvedev%
}{%
{\protect \APACyear {2013}}%
}]{%
Yigit.Medvedev2013_ExtendingParameterization}
\APACinsertmetastar {%
Yigit.Medvedev2013_ExtendingParameterization}%
\begin{APACrefauthors}%
Yi{\u g}it, E.%
\BCBT {}\ \BBA {} Medvedev, A\BPBI S.%
\end{APACrefauthors}%
\unskip\
\newblock
\APACrefYearMonthDay{2013}{}{}.
\newblock
{\BBOQ}\APACrefatitle {Extending the {{Parameterization}} of {{Gravity Waves}}
  into the {{Thermosphere}} and {{Modeling Their Effects}}} {Extending the
  {{Parameterization}} of {{Gravity Waves}} into the {{Thermosphere}} and
  {{Modeling Their Effects}}}.{\BBCQ}
\newblock
\BIn{} F\BHBI J.~L{\"u}bken\ (\BED), \APACrefbtitle {Climate and {{Weather}} of
  the {{Sun-Earth System}} ({{CAWSES}})} {Climate and {{Weather}} of the
  {{Sun-Earth System}} ({{CAWSES}})}\ (\BPGS\ 467--480).
\newblock
\APACaddressPublisher{Dordrecht}{Springer Netherlands}.
\newblock
\begin{APACrefDOI} \doi{10.1007/978-94-007-4348-9_25} \end{APACrefDOI}
\PrintBackRefs{\CurrentBib}

\bibitem [\protect \citeauthoryear {%
Yi{\u g}it%
\ \BBA {} Medvedev%
}{%
Yi{\u g}it%
\ \BBA {} Medvedev%
}{%
{\protect \APACyear {2015}}%
}]{%
Yigit.Medvedev2015_InternalWave}
\APACinsertmetastar {%
Yigit.Medvedev2015_InternalWave}%
\begin{APACrefauthors}%
Yi{\u g}it, E.%
\BCBT {}\ \BBA {} Medvedev, A\BPBI S.%
\end{APACrefauthors}%
\unskip\
\newblock
\APACrefYearMonthDay{2015}{}{}.
\newblock
{\BBOQ}\APACrefatitle {Internal Wave Coupling Processes in {{Earth}}'s
  Atmosphere} {Internal wave coupling processes in {{Earth}}'s
  atmosphere}.{\BBCQ}
\newblock
\APACjournalVolNumPages{Advances in Space Research}{55}{4}{983--1003}.
\newblock
\begin{APACrefDOI} \doi{10.1016/j.asr.2014.11.020} \end{APACrefDOI}
\PrintBackRefs{\CurrentBib}

\bibitem [\protect \citeauthoryear {%
Yi{\u g}it%
\ \BBA {} Medvedev%
}{%
Yi{\u g}it%
\ \BBA {} Medvedev%
}{%
{\protect \APACyear {2019}}%
}]{%
Yigit.Medvedev2019_ObscureWaves}
\APACinsertmetastar {%
Yigit.Medvedev2019_ObscureWaves}%
\begin{APACrefauthors}%
Yi{\u g}it, E.%
\BCBT {}\ \BBA {} Medvedev, A\BPBI S.%
\end{APACrefauthors}%
\unskip\
\newblock
\APACrefYearMonthDay{2019}{}{}.
\newblock
{\BBOQ}\APACrefatitle {Obscure Waves in Planetary Atmospheres} {Obscure waves
  in planetary atmospheres}.{\BBCQ}
\newblock
\APACjournalVolNumPages{Physics Today}{72}{6}{40--46}.
\newblock
\begin{APACrefDOI} \doi{10.1063/PT.3.4226} \end{APACrefDOI}
\PrintBackRefs{\CurrentBib}

\bibitem [\protect \citeauthoryear {%
Yi{\u g}it%
, Medvedev%
, Aylward%
, Hartogh%
\BCBL {}\ \BBA {} Harris%
}{%
Yi{\u g}it%
\ \protect \BOthers {.}}{%
{\protect \APACyear {2009}}%
}]{%
Yigit.etal2009_ModelingEffects}
\APACinsertmetastar {%
Yigit.etal2009_ModelingEffects}%
\begin{APACrefauthors}%
Yi{\u g}it, E.%
, Medvedev, A\BPBI S.%
, Aylward, A\BPBI D.%
, Hartogh, P.%
\BCBL {}\ \BBA {} Harris, M\BPBI J.%
\end{APACrefauthors}%
\unskip\
\newblock
\APACrefYearMonthDay{2009}{}{}.
\newblock
{\BBOQ}\APACrefatitle {Modeling the Effects of Gravity Wave Momentum Deposition
  on the General Circulation above the Turbopause} {Modeling the effects of
  gravity wave momentum deposition on the general circulation above the
  turbopause}.{\BBCQ}
\newblock
\APACjournalVolNumPages{J. Geophys. Res.}{114}{D7}{D07101}.
\newblock
\begin{APACrefDOI} \doi{10.1029/2008JD011132} \end{APACrefDOI}
\PrintBackRefs{\CurrentBib}

\bibitem [\protect \citeauthoryear {%
Yi{\u g}it%
\ \protect \BOthers {.}}{%
Yi{\u g}it%
\ \protect \BOthers {.}}{%
{\protect \APACyear {2012}}%
}]{%
Yigit.etal2012_DynamicalEffects}
\APACinsertmetastar {%
Yigit.etal2012_DynamicalEffects}%
\begin{APACrefauthors}%
Yi{\u g}it, E.%
, Medvedev, A\BPBI S.%
, Aylward, A\BPBI D.%
, Ridley, A\BPBI J.%
, Harris, M\BPBI J.%
, Moldwin, M\BPBI B.%
\BCBL {}\ \BBA {} Hartogh, P.%
\end{APACrefauthors}%
\unskip\
\newblock
\APACrefYearMonthDay{2012}{}{}.
\newblock
{\BBOQ}\APACrefatitle {Dynamical Effects of Internal Gravity Waves in the
  Equinoctial Thermosphere} {Dynamical effects of internal gravity waves in the
  equinoctial thermosphere}.{\BBCQ}
\newblock
\APACjournalVolNumPages{Journal of Atmospheric and Solar-Terrestrial
  Physics}{90--91}{}{104--116}.
\newblock
\begin{APACrefDOI} \doi{10.1016/j.jastp.2011.11.014} \end{APACrefDOI}
\PrintBackRefs{\CurrentBib}

\bibitem [\protect \citeauthoryear {%
Yi{\u g}it%
, Medvedev%
, England%
\BCBL {}\ \BBA {} Immel%
}{%
Yi{\u g}it%
\ \protect \BOthers {.}}{%
{\protect \APACyear {2014}}%
}]{%
Yigit.etal2014_SimulatedVariability}
\APACinsertmetastar {%
Yigit.etal2014_SimulatedVariability}%
\begin{APACrefauthors}%
Yi{\u g}it, E.%
, Medvedev, A\BPBI S.%
, England, S\BPBI L.%
\BCBL {}\ \BBA {} Immel, T\BPBI J.%
\end{APACrefauthors}%
\unskip\
\newblock
\APACrefYearMonthDay{2014}{}{}.
\newblock
{\BBOQ}\APACrefatitle {Simulated Variability of the High-Latitude Thermosphere
  Induced by Small-Scale Gravity Waves during a Sudden Stratospheric Warming}
  {Simulated variability of the high-latitude thermosphere induced by
  small-scale gravity waves during a sudden stratospheric warming}.{\BBCQ}
\newblock
\APACjournalVolNumPages{J. Geophys. Res. Space Phys.}{119}{1}{357--365}.
\newblock
\begin{APACrefDOI} \doi{10.1002/2013JA019283} \end{APACrefDOI}
\PrintBackRefs{\CurrentBib}

\bibitem [\protect \citeauthoryear {%
Yi{\u g}it%
, Medvedev%
\BCBL {}\ \BBA {} Ern%
}{%
Yi{\u g}it%
, Medvedev%
\BCBL {}\ \BBA {} Ern%
}{%
{\protect \APACyear {2021}}%
}]{%
Yigit.etal2021_EffectsLatitudeDependent}
\APACinsertmetastar {%
Yigit.etal2021_EffectsLatitudeDependent}%
\begin{APACrefauthors}%
Yi{\u g}it, E.%
, Medvedev, A\BPBI S.%
\BCBL {}\ \BBA {} Ern, M.%
\end{APACrefauthors}%
\unskip\
\newblock
\APACrefYearMonthDay{2021}{}{}.
\newblock
{\BBOQ}\APACrefatitle {Effects of {{Latitude-Dependent Gravity Wave Source
  Variations}} on the {{Middle}} and {{Upper Atmosphere}}} {Effects of
  {{Latitude-Dependent Gravity Wave Source Variations}} on the {{Middle}} and
  {{Upper Atmosphere}}}.{\BBCQ}
\newblock
\APACjournalVolNumPages{Front. Astron. Space Sci.}{7}{}{614018}.
\newblock
\begin{APACrefDOI} \doi{10.3389/fspas.2020.614018} \end{APACrefDOI}
\PrintBackRefs{\CurrentBib}

\bibitem [\protect \citeauthoryear {%
Yi{\u g}it%
, Medvedev%
, Gann%
, Klaassen%
\BCBL {}\ \BBA {} Rowland%
}{%
Yi{\u g}it%
\ \protect \BOthers {.}}{%
{\protect \APACyear {2025}}%
}]{%
Yigit.etal2025_ImpactGravity}
\APACinsertmetastar {%
Yigit.etal2025_ImpactGravity}%
\begin{APACrefauthors}%
Yi{\u g}it, E.%
, Medvedev, A\BPBI S.%
, Gann, A\BPBI L\BPBI S.%
, Klaassen, G\BPBI P.%
\BCBL {}\ \BBA {} Rowland, D\BPBI E.%
\end{APACrefauthors}%
\unskip\
\newblock
\APACrefYearMonthDay{2025}{}{}.
\newblock
{\BBOQ}\APACrefatitle {Impact of {{Gravity Waves From Tropospheric}} and
  {{Non-Tropospheric Sources}} on the {{Middle}} and {{Upper Atmosphere}} and
  {{Comparison With ICON}}/{{MIGHTI Winds}}} {Impact of {{Gravity Waves From
  Tropospheric}} and {{Non-Tropospheric Sources}} on the {{Middle}} and {{Upper
  Atmosphere}} and {{Comparison With ICON}}/{{MIGHTI Winds}}}.{\BBCQ}
\newblock
\APACjournalVolNumPages{J. Geophys. Res. Space Phys.}{130}{9}{e2025JA034204}.
\newblock
\begin{APACrefDOI} \doi{10.1029/2025JA034204} \end{APACrefDOI}
\PrintBackRefs{\CurrentBib}

\bibitem [\protect \citeauthoryear {%
Yi{\u g}it%
, Medvedev%
\BCBL {}\ \BBA {} Hartogh%
}{%
Yi{\u g}it%
\ \protect \BOthers {.}}{%
{\protect \APACyear {2018}}%
}]{%
Yigit.etal2018_InfluenceGravity}
\APACinsertmetastar {%
Yigit.etal2018_InfluenceGravity}%
\begin{APACrefauthors}%
Yi{\u g}it, E.%
, Medvedev, A\BPBI S.%
\BCBL {}\ \BBA {} Hartogh, P.%
\end{APACrefauthors}%
\unskip\
\newblock
\APACrefYearMonthDay{2018}{}{}.
\newblock
{\BBOQ}\APACrefatitle {Influence of Gravity Waves on the Climatology of
  High-Altitude {{Martian}} Carbon Dioxide Ice Clouds} {Influence of gravity
  waves on the climatology of high-altitude {{Martian}} carbon dioxide ice
  clouds}.{\BBCQ}
\newblock
\APACjournalVolNumPages{Ann. Geophys.}{36}{6}{1631--1646}.
\newblock
\begin{APACrefDOI} \doi{10.5194/angeo-36-1631-2018} \end{APACrefDOI}
\PrintBackRefs{\CurrentBib}

\bibitem [\protect \citeauthoryear {%
Yi{\u g}it%
, Medvedev%
\BCBL {}\ \BBA {} Hartogh%
}{%
Yi{\u g}it%
, Medvedev%
\BCBL {}\ \BBA {} Hartogh%
}{%
{\protect \APACyear {2021}}%
}]{%
Yigit.etal2021_VariationsMartian}
\APACinsertmetastar {%
Yigit.etal2021_VariationsMartian}%
\begin{APACrefauthors}%
Yi{\u g}it, E.%
, Medvedev, A\BPBI S.%
\BCBL {}\ \BBA {} Hartogh, P.%
\end{APACrefauthors}%
\unskip\
\newblock
\APACrefYearMonthDay{2021}{}{}.
\newblock
{\BBOQ}\APACrefatitle {Variations of the {{Martian Thermospheric Gravity-wave
  Activity}} during the {{Recent Solar Minimum}} as {{Observed}} by {{MAVEN}}}
  {Variations of the {{Martian Thermospheric Gravity-wave Activity}} during the
  {{Recent Solar Minimum}} as {{Observed}} by {{MAVEN}}}.{\BBCQ}
\newblock
\APACjournalVolNumPages{ApJ}{920}{2}{69}.
\newblock
\begin{APACrefDOI} \doi{10.3847/1538-4357/ac15fc} \end{APACrefDOI}
\PrintBackRefs{\CurrentBib}

\bibitem [\protect \citeauthoryear {%
Zalucha%
, Brecht%
, Rafkin%
, Bougher%
\BCBL {}\ \BBA {} Alexander%
}{%
Zalucha%
\ \protect \BOthers {.}}{%
{\protect \APACyear {2013}}%
}]{%
Zalucha.etal2013_IncorporationGravity}
\APACinsertmetastar {%
Zalucha.etal2013_IncorporationGravity}%
\begin{APACrefauthors}%
Zalucha, A\BPBI M.%
, Brecht, A\BPBI S.%
, Rafkin, S.%
, Bougher, S\BPBI W.%
\BCBL {}\ \BBA {} Alexander, M\BPBI J.%
\end{APACrefauthors}%
\unskip\
\newblock
\APACrefYearMonthDay{2013}{}{}.
\newblock
{\BBOQ}\APACrefatitle {Incorporation of a Gravity Wave Momentum Deposition
  Parameterization into the {{Venus Thermosphere General Circulation Model}}
  ({{VTGCM}})} {Incorporation of a gravity wave momentum deposition
  parameterization into the {{Venus Thermosphere General Circulation Model}}
  ({{VTGCM}})}.{\BBCQ}
\newblock
\APACjournalVolNumPages{J. Geophys. Res. Planets}{118}{1}{147--160}.
\newblock
\begin{APACrefDOI} \doi{10.1029/2012JE004168} \end{APACrefDOI}
\PrintBackRefs{\CurrentBib}

\end{thebibliography}

%
%

\newpage



\end{document}